\documentclass[aps,twocolumn,prb]{revtex4}%
\usepackage{amsfonts}
\usepackage{amsmath}
\usepackage{amssymb}
\usepackage{natbib}
\usepackage{graphicx}%
\begin{document}
\title[Pseudogap and high T$_{c}$]{Pseudogap and high-temperature superconductivity from weak to strong coupling.
Towards quantitative theory.}
\author{A.-M.S. Tremblay, B. Kyung, D. S\'{e}n\'{e}chal}
\affiliation{D\'{e}partement de physique and RQMP, Universit\'{e} de Sherbrooke,
Sherbrooke, QC J1K 2R1, Canada}
\pacs{PACS number}

\begin{abstract}
This is a short review of the theoretical work on the two-dimensional Hubbard
model performed in Sherbrooke in the last few years. It is written on the occasion of the twentieth
anniversary of the discovery of high-temperature superconductivity. We discuss
several approaches, how they were benchmarked and how they agree sufficiently
with each other that we can trust that the results are accurate solutions of
the Hubbard model. Then comparisons are made with experiment. We show that the
Hubbard model does exhibit d-wave superconductivity and antiferromagnetism
essentially where they are observed for both hole and electron-doped cuprates.
We also show that the pseudogap phenomenon comes out of these calculations. In
the case of electron-doped high temperature superconductors, comparisons with
angle-resolved photoemission experiments are nearly quantitative. The value of
the pseudogap temperature observed for these compounds in recent photoemission
experiments has been predicted by theory before it was observed
experimentally. Additional experimental confirmation would be useful. The
theoretical methods that are surveyed include mostly the Two-Particle
Self-Consistent Approach, Variational Cluster Perturbation Theory (or
variational cluster approximation), and Cellular Dynamical Mean-Field Theory.

\end{abstract}
\date{November 2005}
\maketitle
\tableofcontents

\section{Introduction}

In the first days of the discovery of high-temperature superconductivity,
Anderson\cite{Anderson:1987} suggested that the two-dimensional Hubbard model
held the key to the phenomenon. Despite its apparent simplicity, the
two-dimensional Hubbard model is a formidable challenge for theorists. The
dimension is not low enough that an exact solution is available, as in
one dimension. The dimension is not high enough that some mean-field theory,
like Dynamical Mean Field Theory\cite{Georges:1996, Jarrell:1992} (DMFT), valid in infinite
dimension, can come to the rescue. In two dimensions, both quantum and thermal
fluctuations are important. In addition, as we shall see, it turns out that
the real materials are in a situation where both potential and kinetic energy
are comparable. We cannot begin with the wave picture (kinetic energy
dominated, or so-called \textquotedblleft weak coupling\textquotedblright) and
do perturbation theory, and we cannot begin from the particle picture
(potential energy dominated, or so-called \textquotedblleft strong
coupling\textquotedblright) and do perturbation theory. In fact, even if one
starts from the wave picture, perturbation theory is not trivial in two
dimensions, as we shall see. Variational approaches on the ground state have
been proposed,\cite{Paramekanti:2004} but even if they capture key aspects of
the ground state, they say little about one-particle excitations.

\begin{figure}[ptb]
\centerline{\includegraphics[width=7.5cm]{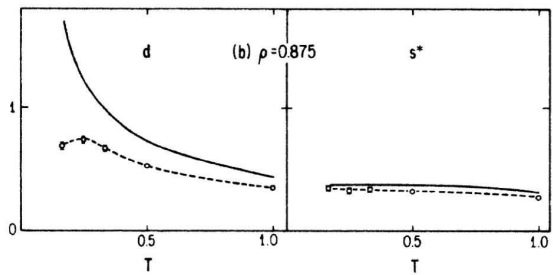}} \caption{ The $d_{x^{2}-y^{2}}$
and extended $s$-wave susceptibilities obtained from QMC simulations for
$U=4t$ and a $4\times4$ lattice. The solid lines are the non-interacting
results. From Ref.~\onlinecite{Hirsch:1988}. The low temperature downturn of d-wave however seems to come from a mistreatment of the sign problem (D.J. Scalapino private communication).}%
\label{fig_0}%
\end{figure}

Even before the discovery of high-temperature superconductivity, it was
suggested that antiferromagnetic fluctuations present in the Hubbard model
could lead to d-wave superconductivity,\cite{Scalapino:1986, Beal-Monod:1986,
Miyake:1986} a sort of generalization of the Kohn-Luttinger
mechanism\cite{Kohn:1965} analogous to the superfluidity mediated by
ferromagnetic spin fluctuations in $^{3}$He.\cite{Leggett:1975} Nevertheless,
early Quantum Monte Carlo (QMC) simulations\cite{Hirsch:1988} gave rather
discouraging results, as illustrated in Fig.~\ref{fig_0}. In QMC, low
temperatures are inaccessible because of the sign problem. At accessible
temperatures, the d-wave pair susceptibility is smaller than the
non-interacting one, instead of diverging. Since the observed phenomenon
appears at temperatures that are about ten times smaller than what is
accessible with QMC, the problem was left open.
Detailed analysis of the irreducible vertex\cite{Bulut:1993} deduced from QMC did suggest the importance of d-wave pairing, but other numerical work\cite{Zhang:1997} concluded that long-range d-wave order is absent, despite the fact that slave-boson
approaches\cite{Kotliar:1988, Inui:1988} and many subsequent work suggested otherwise. The situation on the numerical side is changing since
more recent
variational,\cite{Paramekanti:2004} Dynamical Cluster
Approximation\cite{Maier:2000a} and exact diagonalization\cite{Poilblanc:2002}
results now point towards the existence of d-wave superconductivity in the
Hubbard model. Even more recently, new numerical approaches are making an even
more convincing case.\cite{Senechal:2005, Kancharla:2005, Maier:2005a}

After twenty years, we should be as quantitative as possible.
How should we proceed to investigate a model without a small parameter?
We will try to follow this path: (1) Identify important physical principles and laws to constrain
non-perturbative approximation schemes, starting from both weak (kinetic
energy dominated) and strong (potential energy dominated) coupling. (2)
Benchmark the various approaches as much as possible against exact (or
numerically accurate) results. (3) Check that weak and strong coupling
approaches agree at intermediate coupling. (4) Compare with experiment.

In brief, we are trying to answer the question, \textquotedblleft Is the
Hubbard model rich enough to contain the essential physics of the cuprates,
both hole and electron doped?"
The answer is made possible by new theoretical approaches, increased
computing power, and the reassurance that theoretical approaches, numerical
and analytical, give consistent results at intermediate coupling even if the
starting points are very different.

This paper is a review of the work we have done in Sherbrooke on this subject.
In the short space provided, this review will not cover all of our work.
Needless to say, we will be unfair to the work of many other groups, even
though we will try to refer to the work of others that is directly relevant to
ours. We do not wish to make priority claims and we apologize to the authors
that may feel unfairly treated.

Section \ref{Methodology} will introduce the methodology: First a method that
is valid at weak to intermediate coupling, the Two-Particle Self-Consistent
approach (TPSC), and then various quantum cluster methods that are better at
strong coupling, namely Cluster Perturbation Theory (CPT), the Variational
Cluster Approximation (VCA) also known as Variational Cluster Perturbation
Theory (VCPT), and Cellular Dynamical Mean Field Theory (CDMFT) with a brief
mention of the Dynamical Cluster Approximation (DCA). In all cases, we will
mention the main comparisons with exact or numerically accurate results that
have been used to benchmark the approaches. In Sect.~\ref{Results} we give
some of the results, mostly on the pseudogap and the phase diagram of
high-temperature superconductors. More importantly perhaps, we show the
consistency of the results obtained by both weak- and strong-coupling
approaches when they are used at intermediate coupling. Finally, we compare
with experiment in section \ref{Experiment}.

\section{Methodology\label{Methodology}}

We consider the Hubbard model%
\begin{equation}
H=-\sum_{i,j,\sigma}t_{ij}c_{i\sigma}^{\dagger}c_{j\sigma}+U\sum
_{i}n_{i\uparrow}n_{i\downarrow}%
\end{equation}
where $c_{i\sigma}^{\dagger}$ ($c_{i\sigma}$) are creation and annihilation
operators for electrons of spin $\sigma$, $n_{i\sigma}=c_{i\sigma}^{\dagger
}c_{i\sigma}$ is the density of spin $\sigma$ electrons, $t_{ij}=t_{ji}^{\ast
}$ is the hopping amplitude, and $U$ is the on-site Coulomb repulsion. In
general, we write $t,t^{\prime},t^{\prime\prime}$ respectively for the first-,
second- and third-nearest neighbor hopping amplitudes.

In the following subsections, we first discuss how to approach the problem
from the weak coupling perspective and then from the strong coupling point of
view. The approaches that we will use in the end are non-perturbative, but in
general they are more accurate either at weak or strong coupling.

\subsection{Weak coupling approach}

Even at weak coupling, the Hubbard model presents difficulties specific to two dimensions.
The time-honored Random Phase Approximation (RPA) has the
advantage of satisfying conservation laws, but it violates the Pauli principle
and the Mermin-Wagner-Hohenberg-Coleman (or Mermin-Wagner, for short) theorem.
This theorem states that a continuous symmetry cannot be broken at finite
temperature in two dimensions. RPA gives a finite-temperature phase
transition. The Pauli principle means, in particular, that $\langle
n_{i\uparrow}n_{i\uparrow}\rangle=\langle n_{i\uparrow}\rangle$ in a model
with only one orbital per site. This is violated by RPA since it can be
satisfied only if all possible exchanges of electron lines are allowed (more
on this in the following section). Since the square of the density at a given
site is given by $\langle(n_{i\uparrow}+n_{i\downarrow})^{2}\rangle=2\langle
n_{\uparrow}n_{\uparrow}\rangle+2\langle n_{\uparrow}n_{\downarrow}\rangle$,
violating the Pauli condition $\langle n_{i\uparrow}n_{i\uparrow}%
\rangle=\langle n_{i\uparrow}\rangle$ will in general lead to large errors in
double occupancy, a key quantity in the Hubbard model since it is proportional
to the potential energy. Another popular approach is the
Moriya\cite{Moriya:1985} self-consistent spin-fluctuation
approach\cite{Markiewicz:2004} that uses a Hubbard-Stratonovich transformation
and a $\langle\phi^{4}\rangle\sim\phi^{2}\langle\phi^{2}\rangle$
factorization. This satisfies the Mermin-Wagner theorem but, unfortunately,
violates the Pauli principle and introduces an unknown mode-coupling constant
as well as an unknown renormalized $U$ in the second-order term. The
conserving approximation known as Fluctuation Exchange (FLEX)
approximation\cite{Bickers:1989} is an Eliashberg-type theory that is
conserving but violates the Pauli principle, assumes a Migdal theorem and does
not reproduce the pseudogap phenomenon observed in QMC. More detailed
criticism of this and other approaches may be found in
Refs.~\onlinecite{Vilk:1997,Vilk:1996}. Finally, the renormalization
group\cite{Honerkamp:2001, Halboth:2000, Furukawa:1998, Wegner:2003,
Hankevych:2003a} has the great advantage of being an unbiased method to look
for instabilities towards various ordered phases. However, it is quite
difficult to implement in two dimensions because of the proliferation of
coupling constants, and, to our knowledge, no one has yet implemented a
two-loop calculation without introducing additional
approximations.\cite{Zanchi:2001,Katanin:2004} Such a two-loop calculation is
necessary to observe the pseudogap phenomenon.

\subsubsection{Two-Particle Self-consistent approach (TPSC)}

The TPSC approach, originally proposed by Vilk, Tremblay and collaborators,
\cite{Vilk:1994, Vilk:1995} aims at capturing non-perturbative effects. It does not use
perturbation theory or, if you want, it drops diagrammatic expansions.
Instead, it is based on imposing constraints and sum rules: the theory should
satisfy (a) the spin and charge conservation laws (b) the Pauli principle in
the form $\langle n_{i\uparrow}n_{i\uparrow}\rangle=\langle n_{i\uparrow
}\rangle$ (c) the local-moment and the local-density sum rules. Without any
further explicit constraint, we find that the theory satisfies the
Mermin-Wagner theorem, that it satisfies consistency between one- and
two-particle quantities in the sense that $\frac{1}{2}\mathrm{Tr}(\Sigma
G)=U\langle n_{\uparrow}n_{\downarrow}\rangle$ and finally that the theory
contains the physics of Kanamori-Br\"{u}ckner screening (in other words,
scattering between electrons and holes includes T-matrix quantum fluctuation
effects beyond the Born approximation).

Several derivations of our approach have been given,\cite{Vilk:1995,
Mahan:2000} including a quite formal one\cite{Allen:2003} based on the
functional derivative Baym-Kadanoff approach.\cite{Baym:1962} Here we only
give an outline\cite{Moukouri:2000} of the approach with a more
phenomenological outlook. We proceed in two steps. In the first step (in our
earlier work sometimes called zeroth step), the self-energy is
obtained by a Hartree-Fock-type factorization of the four-point function with
the \textit{additional constraint} that the factorization is exact when all
space-time coordinates coincide.\footnote{This additional constraint leads to
a degree of consistency between one- and two-particle quantities that is
absent from the standard Hartree-Fock factorization.} Functional
differentiation, as in the Baym-Kadanoff approach\cite{Baym:1962}, then leads
to a momentum- and frequency-independent irreducible particle-hole vertex for
the spin channel that satisfies\cite{Vilk:1994} $U_{sp}=U\langle n_{\uparrow
}n_{\downarrow}\rangle/(\langle n_{\uparrow}\rangle\langle n_{\downarrow
}\rangle)$. The local moment sum rule and the Pauli principle in the form
$\langle n_{\sigma}^{2}\rangle=\langle n_{\sigma}\rangle$ then determine
double occupancy and $U_{sp}$. The irreducible vertex for the charge channel
is too complicated to be computed exactly, so it is assumed to be constant and
its value is found from the Pauli principle and the local charge fluctuation
sum rule. To be more specific, let us use the notation, $q=(\mathbf{q,}%
iq_{n})$ and $k=(\mathbf{k,}ik_{n})$ with $iq_{n}$ and $ik_{n}$ respectively
bosonic and fermionic Matsubara frequencies. We work in units where $k_{B},$
$\hbar,$ and lattice spacing are all unity. The spin and charge
susceptibilities now take the form
\begin{equation}
\chi_{sp}^{-1}(q)=\chi_{0}(q)^{-1}-\frac12 U_{sp} \label{Chi_sp}%
\end{equation}
and
\begin{equation}
\chi_{ch}^{-1}(q)=\chi_{0}(q)^{-1}+\frac12 U_{ch}%
\end{equation}
with $\chi_{0}$ computed with the Green function $G_{\sigma}^{(1)}$ that
contains the self-energy whose functional differentiation gave the vertices.
This self-energy is constant, corresponding to the Hartree-Fock-type
factorization.\footnote{The constant self-energy is \textit{not} equal to
$Un_{-\sigma}$ as it would in the trivial Hartree-Fock factorization (see
previous note).} The susceptibilities thus satisfy conservation
laws\cite{Baym:1962}. One enforces the Pauli
principle $\langle n_{\sigma}^{2}\rangle=\langle n_{\sigma}\rangle$ implicit
in the following two sum rules,
\begin{align}
\frac{T}{N}\sum_{q}\chi_{sp}(q)  &  =\left\langle (n_{\uparrow}-n_{\downarrow
})^{2}\right\rangle =n-2\langle n_{\uparrow}n_{\downarrow}\rangle
\label{Suscep}\\
\frac{T}{N}\sum_{q}\chi_{ch}(q)  &  =\left\langle (n_{\uparrow}+n_{\downarrow
})^{2}\right\rangle -n^{2}=n+2\langle n_{\uparrow}n_{\downarrow}\rangle
-n^{2}\nonumber
\end{align}
where $n$ is the density. The above equations, in addition to\cite{Vilk:1994}
\begin{equation}
U_{sp}=\frac{U\langle n_{\uparrow}n_{\downarrow}\rangle}{\langle n_{\uparrow
}\rangle\langle n_{\downarrow}\rangle}, \label{ansatz}%
\end{equation}
suffice to determine the constant vertices $U_{sp}$ and $U_{ch}$.

Once the two-particle quantities have been found as above, the next step of
the approach of Ref.~\onlinecite{Vilk:1997}, consists in improving the approximation
for the single-particle self-energy by starting from an exact expression where
the high-frequency Hartree-Fock behavior is explicitly factored out. One then
substitutes in the exact expression the irreducible low-frequency vertices
$U_{sp}$ and $U_{ch}$ as well as $G_{\sigma}^{(1)}(k+q)$ and $\chi
_{sp}(q),\chi_{ch}(q)$ computed above. The exact form for the self-energy
expression can however be obtained either in the longitudinal or in the
transverse channel. To satisfy crossing symmetry of the fully reducible vertex
appearing in the general expression and to preserve consistency between one-
and two-particle quantities, one averages the two possibilities to obtain
\cite{Moukouri:2000}%
\begin{align}
&  \Sigma_{\sigma}^{(2)}(k)=Un_{-\sigma}\nonumber\label{Self-long}\\
&  ~~+\frac{U}{8}\frac{T}{N}\sum_{q}\left[3U_{sp}\chi_{sp}(q)+U_{ch}%
\chi_{ch}(q)\right]  G_{\sigma}^{(1)}(k+q).
\end{align}
The resulting self-energy $\Sigma_{\sigma}^{(2)}(k)$ on the left hand-side is
at the next level of approximation so it differs from the self-energy entering
the right-hand side. One can verify that the longitudinal spin fluctuations
contribute an amount $U\langle n_{\uparrow}n_{\downarrow}\rangle/4$ to the
consistency condition\cite{Vilk:1996} $\frac{1}{2}\mathrm{Tr}(\Sigma
^{(2)}G^{(1)})=$ $U\langle n_{\uparrow}n_{\downarrow}\rangle$ and that each of
the two transverse spin components as well as the charge fluctuations also
each contribute $U\langle n_{\uparrow}n_{\downarrow}\rangle/4.$ In addition,
one verifies numerically that the exact sum rule\cite{Vilk:1997} $-\int
d\omega^{\prime}\operatorname{Im}[\Sigma_{\sigma}(\mathbf{k,}\omega^{\prime
})]/\pi=U^{2}n_{-\sigma}(1-n_{-\sigma})$ determining the high-frequency
behavior is satisfied to a high degree of accuracy.

The theory also has a consistency check. Indeed, the exact expression for
consistency between one- and two-particle quantities should be written with
$G^{(2)}$ given by $(G^{-1})^{(2)}=(G^{-1})^{(0)}-\Sigma^{(2)}$ instead of
with $G^{(1)}$. In other words $\frac{1}{2}\mathrm{Tr}(\Sigma^{(2)}G^{(2)})=$
$U\langle n_{\uparrow}n_{\downarrow}\rangle$ should be satisfied instead of
$\frac{1}{2}\mathrm{Tr}(\Sigma^{(2)}G^{(1)})=$ $U\langle n_{\uparrow
}n_{\downarrow}\rangle,$ which is exactly satisfied here. We find through
QMC benchmarks that when the left- and right-hand side of the last equation differ
only by a few percent, then the theory is accurate.

To obtain the thermodynamics, one finds the entropy by integrating $1/T$ times the
specific heat $(\partial E/\partial T)$ so that we know $F=E-TS$. There are
other ways to obtain the thermodynamics and one looks for consistency between
these.\cite{Roy:unpub} We will not discuss thermodynamic aspects in the
present review.

At weak coupling in the repulsive model the particle-hole channel is the one
that is influenced directly. Correlations in crossed channels, such as pairing
susceptibilities, are induced indirectly and are harder to evaluate. This
simply reflects the fact the simplest Hartree-Fock factorization of the
Hubbard model does not lead to a d-wave order parameter (even though
Hartree-Fock factorization of its strong-coupling version does). The
$d_{x^{2}-y^{2}}$-wave susceptibility is defined by $\chi_{d}=\int_{0}^{\beta
}d\tau\langle T_{\tau}\Delta( \tau) \Delta^{\dagger}\rangle$ with the $d$-wave
order parameter equal to $\Delta^{\dagger}=\sum_{i}\sum_{\gamma}g( \gamma)
c_{i\uparrow}^{\dagger} c_{i+\gamma\downarrow}^{\dagger}$ the sum over
$\gamma$ being over nearest-neighbors, with $g( \gamma) =\pm1/2$ depending on
whether $\gamma$ is a neighbor along the $\hat{\mathbf{x}}$ or the
$\hat{\mathbf{y}}$ axis. Briefly speaking,\cite{Kyung:2003, Allen:unpub}
to extend TPSC to compute pairing susceptibility, we begin from the
Schwinger-Martin-Kadanoff-Baym formalism with both
diagonal\cite{Vilk:1997,Allen:2003} and off-diagonal\cite{Allen:2001} source
fields. The self-energy is expressed in terms of spin and charge fluctuations
and the irreducible vertex entering the Bethe-Salpeter equation for the
pairing susceptibility is obtained from functional differentiation. The final
expression for the $d$-wave susceptibility is,
\begin{widetext}
\begin{align}
\chi_{d}( \mathbf{q}=0,iq_{n}=0)  &  =\frac{T}{N}\sum_{k}\left(  g_{d}%
^{2}(\mathbf{k}) G_{\uparrow}^{(2)}( -k) G_{\downarrow}^{(2)}(k)\right)
-\frac{U}{4}\left(  \frac{T}{N}\right)  ^{2}\sum_{k,k^{\prime}}g_{d}(
\mathbf{k}) G_{\uparrow}^{( 2) }( -k) G_{\downarrow}^{( 2) }( k)\nonumber\\
&  \times\left(  \frac{3}{1-\frac12 U_{sp}\chi_{0}( k^{\prime}-k)} +\frac
{1}{1+\frac12 U_{ch}\chi_{0}(k^{\prime}-k)}\right)  G_{\uparrow}^{( 1) }(
-k^{\prime}) G_{\downarrow}^{(1)}(k^{\prime}) g_{d}(\mathbf{k}^{\prime}).
\label{Suscep_d}%
\end{align}%
\end{widetext}%
In the above expression, $g_{d}( \mathbf{k}) $ is the
form factor for the gap symmetry, while $k$ and $k^{\prime}$ stand for both
wave-vector and fermionic Matsubara frequencies on a square-lattice with $N$
sites at temperature $T.$ The spin and charge susceptibilities take the form
$\chi_{sp}^{-1}( q) =\chi_{0}(q)^{-1}-\frac12 U_{sp}$ and $\chi_{ch}^{-1}(
q) =$ $\chi_{0}(q)^{-1}+\frac12 U_{ch}$ with $\chi_{0}$ computed with the
Green function $G_{\sigma}^{(1)}$ that contains the self-energy whose
functional differentiation gave the spin and charge vertices. The values of
$U_{sp},$ $U_{ch}$ and $\langle n_{\uparrow}n_{\downarrow}\rangle$ are
obtained\ as described above. In the pseudogap regime, one cannot use
$U_{sp}=U\langle n_{\uparrow}n_{\downarrow}\rangle/( \langle n_{\uparrow
}\rangle\langle n_{\downarrow}\rangle)$. Instead,\cite{Vilk:1997} one uses the
local-moment sum rule with the zero temperature value of $\langle n_{\uparrow
}n_{\downarrow}\rangle$ obtained by the method of Ref.~\onlinecite{Lilly:1990}
that agrees very well with QMC calculations at all values of $U.$ Also,
$G_{\sigma}^{(2)}$ contains self-energy effects coming from spin and charge
fluctuations, as described above.\cite{Moukouri:2000, Allen:2003}

The same principles and methodology can be applied for the attractive Hubbard
model.\cite{Kyung:2001, Allen:2001, Allen:unpub} In that case, the dominant
channel is the s-wave pairing channel. Correlations in the crossed channel,
namely the spin and charge susceptibilities, can also be obtained
\textit{mutatis mutandi }along the lines of the previous paragraph.

\subsubsection{Benchmarks for TPSC}

To test any non-perturbative approach, we need reliable benchmarks. Quantum
Monte Carlo (QMC) simulations provide such benchmarks. The results of such numerical
calculations are unbiased and they can be obtained on much larger system sizes
than any other simulation method. The statistical uncertainty can be made as
small as required. The drawback of QMC is that the sign problem renders
calculations impossible at temperatures low enough to reach those that are
relevant for d-wave superconductivity. Nevertheless, QMC can be performed in
regimes that are non-trivial enough to allow us to eliminate some theories on
the grounds that they give qualitatively incorrect results. Comparisons with
QMC allow us to estimate the accuracy of the theory. An approach like TPSC can
then be extended to regimes where QMC is unavailable with the confidence
provided by agreement between both approaches in regimes where both can be performed.

\begin{figure}[ptb]
\centerline{\includegraphics[width=5.0cm]{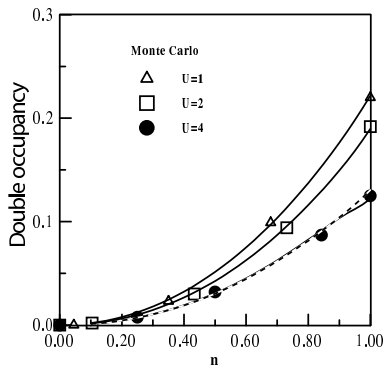}} \caption{Comparisons between
the QMC simulations (symbols) and TPSC (solid lines) for the filling
dependence of the double occupancy. The results are for $T=t/6$ as a function
of filling and for various values of $U$ expect for $U=4t$ where the dashed
line shows the results of our theory at the crossover temperature $T=T_{X}$.
From Ref.~\onlinecite{Vilk:1997}.}%
\label{fig_25}%
\end{figure}

\begin{figure}[ptb]
\centerline{\includegraphics[width=7cm]{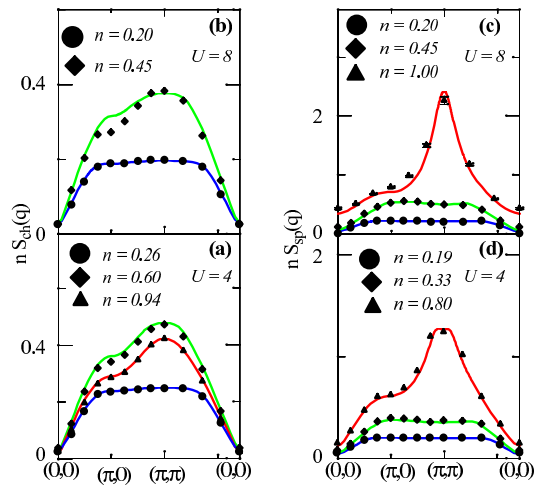}}
\caption{Wave vector ($\mathbf{q}$) dependence of the spin and charge structure
factors for different sets of parameters. Solid lines are from TPSC and
symbols are our QMC data. Monte Carlo data for $n=1$ and $U=8t$ are
for $6 \times6$ clusters and $T=0.5t$; all other data are for $8 \times8$
clusters and $T=0.2t$. Error bars are shown only when significant. From
Ref.~\onlinecite{Vilk:1994}.}%
\label{fig_26}%
\end{figure}

\begin{figure}[ptb]
\centerline{\includegraphics[width=5.5cm]{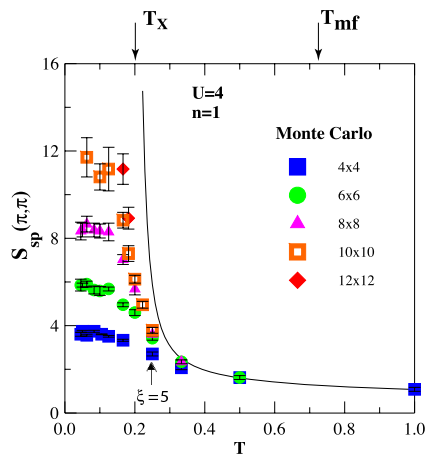}} \caption{Temperature
dependence of $S_{sp}(\pi,\pi)$ at half-filling $n=1$. The solid line is from
TPSC and symbols are Monte Carlo data from Ref.~\onlinecite{White:1989}. Taken
from Ref.~\onlinecite{Vilk:1994}.}%
\label{fig_27}%
\end{figure}

\begin{figure}[ptb]
\centerline{\includegraphics[width=5.5cm]{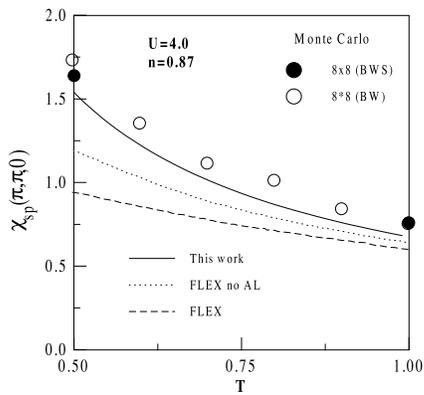}} \caption{Comparisons between
Monte Carlo simulations (BW), FLEX calculations and TPSC for the spin
susceptibility at $Q=(\pi,\pi)$ as a function of temperature at zero Matsubara
frequency. The filled circles (BWS) are from Ref.~\onlinecite{Bulut:1993}.
Taken from Ref.~\onlinecite{Vilk:1997}.}%
\label{fig_28}%
\end{figure}

In order to be concise, details are left to figure captions. Let us first
focus on quantities related to spin and charge fluctuations. The symbols on
the figures refer to QMC results while the solid lines come from TPSC
calculations. Fig.~\ref{fig_25} shows double occupancy, a quantity that plays
a very important role in the Hubbard model in general and in TPSC in
particular. That quantity is shown as a function of filling for various values
of $U$ at inverse temperature $\beta=6$. Fig.~\ref{fig_26} displays the spin
and charge structure factors in a regime where size effects are not important.
Clearly the results are non-perturbative given the large difference between
the spin and charge structure factors, which are plotted here in units where
they are equal at $U=0$. In Fig.~\ref{fig_27} we exhibit the static structure
factor at half-filling as a function of temperature. Below the crossover
temperature $T_{X}$, there is an important size dependence in the QMC results.
The TPSC calculation, represented by a solid line, is done for an infinite
system. We see that the mean-field finite transition temperature $T_{MF}$ is
replaced by a crossover temperature $T_{X}$ at which the correlations enter an
exponential growth regime. One can show analytically\cite{Vilk:1994,
Vilk:1997} that the correlation length becomes infinite only at zero
temperature, thus satisfying the Mermin-Wagner theorem.
The QMC results approach the TPSC results as the system size
grows. Nevertheless, TPSC is in the $N=\infty$ universality
class\cite{Dare:1996} contrary to the Hubbard model for which $N=3,$ so one
expects quantitative differences to increase as the correlation length becomes
larger. It is important to note that $T_{X}$ does not coincide with the
mean-field transition temperature $T_{MF}$. This is because of
Kanamori-Brueckner screening\cite{Chen:1991, Vilk:1994} that manifests itself
in the difference between $U_{sp}$ and the bare $U$. Below $T_{X}$, the main
contribution to the static spin structure factor in Fig.~\ref{fig_27} comes
from the zero-Matsubara frequency component of the spin susceptibility. This
is the so-called renormalized classical regime where the characteristic spin
fluctuation frequency $\omega_{sp}$ is much less than temperature. Even at
temperatures higher than that, TPSC agrees with QMC calculation much better
than other methods, as shown in Fig.~\ref{fig_28}.

\begin{figure}[ptb]
\centerline{\includegraphics[width=6.5cm]{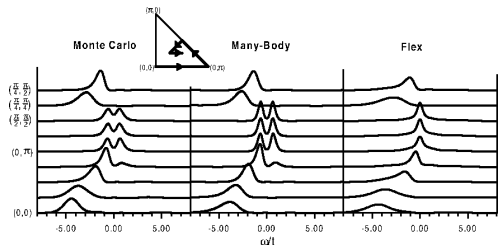}}
\caption{Single particle spectral weight $A(\mathbf{k},\omega)$ for $U=4t$,
$\beta=5/t$, $n=1$, and all independent wave vectors $\mathbf{k}$ of an $8
\times8$ lattice. Results obtained from Maximum Entropy inversion of QMC data
on the left panel and many-body TPSC calculations with Eq.(~\ref{Self-long}) on the
middle panel and with FLEX on the right panel. From
Ref.~\onlinecite{Moukouri:2000}.}%
\label{fig_29}%
%
\end{figure}

\begin{figure}[ptb]
\centerline{\includegraphics[width=7.5cm]{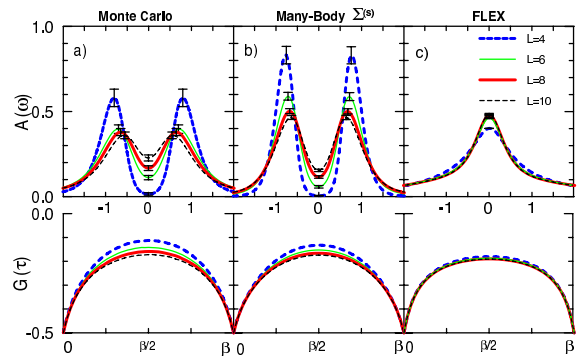}}
\caption{Size dependent
results for various types of calculations for $U=4t$, $\beta=5/t$, $n=1$,
$\mathbf{k}=(0,\pi)$, $L=4,6,8,10$. Upper panels show $A(\mathbf{k},\omega)$
extracted from Maximum Entropy on $G(\tau)$ shown on the corresponding lower
panels. (a) QMC. (b) TPSC using Eq.~(\ref{Self-long}). (c) FLEX. From
Ref.~\onlinecite{Moukouri:2000}.}%
\label{fig_30}%
\end{figure}

Below the crossover temperature to the renormalized classical regime, a
pseudogap develops in the single-particle spectral weight. This is illustrated
in Fig.~\ref{fig_29}.\cite{Moukouri:2000} Eliashberg-type approaches such as
FLEX do not show the pseudogap present in QMC. The size dependence of the
results is also quite close in TPSC and in QMC, as shown in Fig.~\ref{fig_30}.

\begin{figure}[ptb]
\centerline{\includegraphics[width=6.5cm]{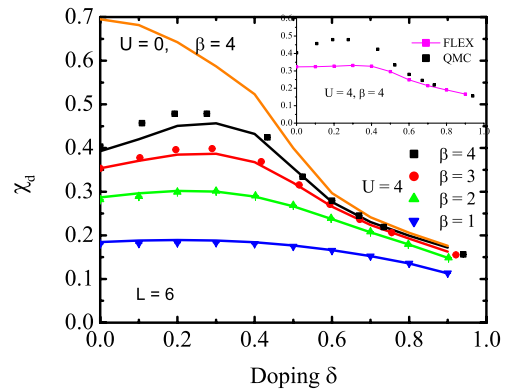}}\caption{Comparisons between
the $d_{x^{2}-y^{2}}$ susceptibility obtained from QMC simulations (symbols)
and from the TPSC approach (lines) in the two-dimensional Hubbard model. Both
calculations are for $U=4t$, a $6\times6$ lattice. QMC error bars are smaller
than the symbols. Analytical results are joined by solid lines. The size
dependence of the results is small at these temperatures. The $U=0$ case is
also shown at $\beta=4/t$ as the upper line. The inset compares QMC and FLEX
at $U=4,\beta=4/t$. From Ref.~\onlinecite{Kyung:2003}.}%
\label{fig_12}%
\end{figure}

The d-wave susceptibility\cite{Kyung:2003} shown in Fig.~\ref{fig_12} again
clearly demonstrates the agreement between TPSC and QMC. In particular, the
dome shape dependence of the QMC results is reproduced to within a few
percent. We will see in Sec. \ref{Results} how one understands the dome shape
and the fact that the d-wave susceptibility of the interacting system is
smaller than that of the non-interacting one in this temperature range.

\begin{figure}[ptb]
\centerline{\includegraphics[width=8.5cm]{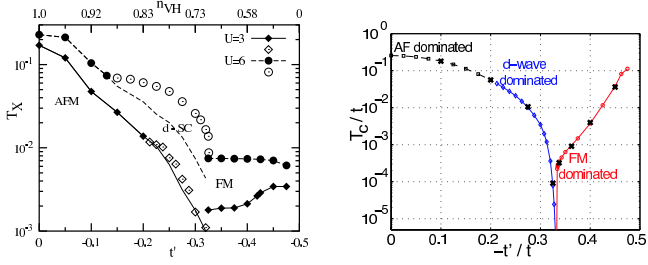}} \caption{The crossover diagram
as a function of next-nearest-neighbor hopping $t^{\prime}$ from TPSC (left)
and from a temperature cutoff renormalization group technique from
Ref.~\onlinecite{Honerkamp:2001} (right). The corresponding Van Hove filling
is indicated on the upper horizontal axis. Crossover lines for magnetic
instabilities near the antiferromagnetic and ferromagnetic wave vectors are
represented by filled symbols while open symbols indicate instability towards
$d_{x^{2}-y^{2}}$-wave superconducting. The solid and dashed lines below the
empty symbols show, respectively for $U=3t$ and $U=6t$, where the
antiferromagnetic crossover temperature would have been in the absence of the
superconducting instability. The largest system size used for this calculation
is $2048\times2048$. From Ref.~\onlinecite{Hankevych:2003}.}%
\label{fig_32}%
\end{figure}
\begin{figure}[ptb]
\centerline{\includegraphics[width=6.0cm]{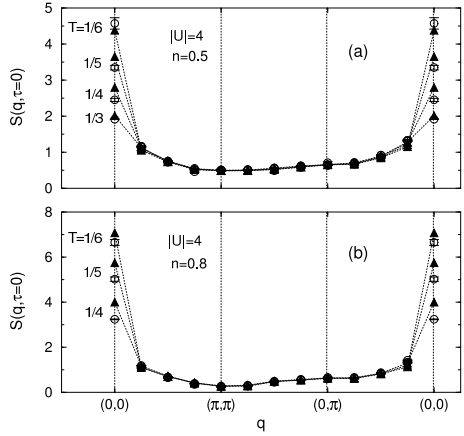}} \caption{TPSC s-wave paring
structure factor $S(\mathbf{q},\tau=0)$ (filled triangles)
and QMC $S(\mathbf{q},\tau=0)$ (open circles) for $U=-4t$ and various temperatures (a) at
$n=0.5$ and (b) at $n=0.8$ on a $8\times8$ lattice. The dashed lines are to
guide the eye. From Ref.~\onlinecite{Kyung:2001}.}%
\label{fig_33}%
\end{figure}
\begin{figure}[ptb]
\centerline{\includegraphics[width=8.5cm]{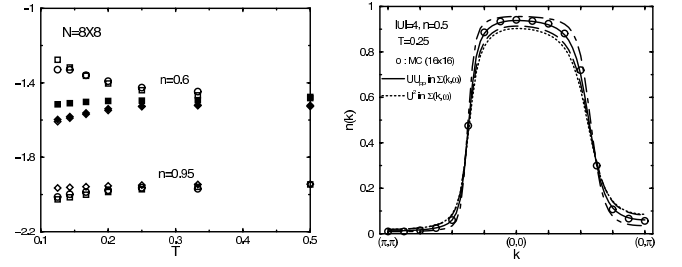}} \caption{Left: chemical
potential shifts $\mu^{(1)}-\mu_{0}$ (open diamonds) and $\mu^{(2)}-\mu_{0}$
(open squares) with the results of QMC calculations (open circles) for
$U=-4t$. Right: The momentum dependent occupation number $n(\mathbf{k})$.
Circles: QMC calculations from Ref.~\onlinecite{Trivedi:1995}. The solid
curve: TPSC. The dashed curve obtained by replacing $U_{pp}$ by $U$ in the
self-energy with all the rest unchanged. The long-dash line is the result of a
self-consistent T-matrix calculation, and the dot-dash line the result of
second-order perturbation theory. From Ref.~\onlinecite{Kyung:2001}.}%
\label{fig_34}%
\end{figure}
\begin{figure}[ptb]
\centerline{\includegraphics[width=8.5cm]{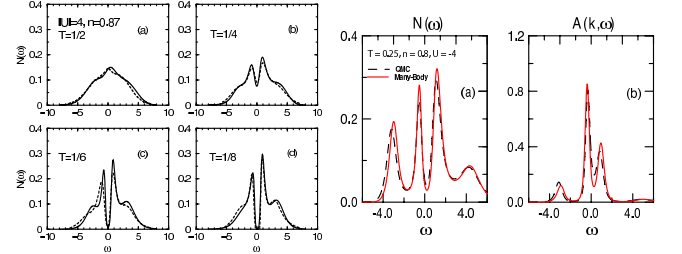}} \caption{Comparisons of local
density of states and single-particle spectral weight from TPSC (solid lines)
and QMC (dashed lines) on a $8\times8$ lattice. QMC data for the density of
states taken from Ref. \onlinecite{Moreo:1992}. Figures from
Ref.~\onlinecite{Kyung:2001}.}%
\label{fig_35}%
\end{figure}

To conclude this section, we quickly mention a few other results obtained with
TPSC. Fig.~\ref{fig_32} contrasts the crossover phase diagram obtained for the
Hubbard model at the van Hove filling\cite{Hankevych:2003} with the results of
a renormalization group calculation.\cite{Honerkamp:2001} The difference
occurring in the ferromagnetic region is discussed in detail in
Ref.~\onlinecite{Hankevych:2003}. Finally, we point out various comparisons
for the attractive Hubbard model. Fig.~\ref{fig_33} shows the static s-wave
pairing susceptibility, Fig.~\ref{fig_34} the chemical potential and the
occupation number, and finally Fig.~\ref{fig_35} the local density of states
and the single-particle spectral weight at a given wave vector.

\subsection{Strong-coupling approaches: Quantum clusters}

DMFT\cite{Georges:1992, Jarrell:1992} has been extremely successful in helping
us understand the Mott transition, the key physical phenomenon that manifests
itself at strong coupling. However, in high dimension where this theory
becomes exact, spatial fluctuations associated with incipient order do not
manifest themselves in the self-energy. In low dimension, this is not the
case. The self-energy has strong momentum dependence, as clearly shown
experimentally in the high-temperature superconductors, and theoretically in
the TPSC approach, a subject we shall discuss again below. It is thus
necessary to go beyond DMFT by studying clusters instead of a single Anderson
impurity as done in DMFT. The simplest cluster approach that includes
strong-coupling effects and momentum dependence is Cluster Perturbation Theory
(CPT).\cite{Gros:1993, Senechal:2000} In this approach, an infinite set of
disconnected clusters are solved exactly and then connected to each other
using strong-coupling perturbation theory. Although the resulting theory turns
out to give the exact result in the $U=0$ case, its derivation clearly shows
that one expects reliable results mostly at strong coupling. This approach
does not include the self-consistent effects contained in DMFT.
Self-consistency or clusters was suggested in Ref. \onlinecite{Georges:1996,Schiller:1995} and a causal approach was first implemented within DCA,\cite{Hettler:1998} where a
momentum-space cluster is connected to a self-consistent momentum-space
medium. In our opinion, the best framework to understand all other cluster
methods is the Self-Energy Functional approach of
Potthoff.\cite{Potthoff:2003a, Potthoff:2003} The form of the lattice Green function obtained
in this approach is the same as that obtained in CPT, clearly exhibiting that
such an approach is better at strong-coupling, even though results often
extrapolate correctly to weak coupling. Amongst the special cases of this
approach, the Variational Cluster Approach (VCA), or Variational Cluster
Perturbation Theory (VCPT)\cite{Potthoff:2003} is the one closest to the
original approach. In a variant, Cellular Dynamical Mean Field
Theory\cite{Kotliar:2001} (CDMFT), a cluster is embedded in a self-consistent
medium instead of a single Anderson impurity as in DMFT (even though the latter approach is accurate in many realistic cases, especially in three dimensions). The strong-coupling
aspects of CDMFT come out clearly in
Refs.~\onlinecite{Stanescu:2004, Stanescu:2005}. A detailed review of quantum
cluster methods has appeared in Ref.~\onlinecite{Maier:2005}.

\subsubsection{Cluster perturbation theory}

Even though CPT does not have the self-consistency present in DMFT type
approaches, at fixed computing resources it allows for the best momentum
resolution. This is particularly important for the ARPES pseudogap in
electron-doped cuprates that has quite a detailed momentum space structure,
and for d-wave superconducting correlations where the zero temperature pair
correlation length may extend well beyond near-neighbor sites. CPT was
developed by Gros\cite{Gros:1993} and S\'{e}n\'{e}chal\cite{Senechal:2000}
independently. This approach can be viewed as the first term of a systematic
expansion around strong coupling.\cite{Senechal:2002} Let us write the hopping
matrix elements in the form
\begin{equation}
t_{\mu\nu}^{mn}=t_{\mu\nu}^{(c)}\delta_{mn}+V_{\mu\nu}^{mn}%
\end{equation}
where $m$ and $n$ label the different clusters, and $\mu,\nu$ label the sites
within a cluster. Then $t_{\mu\nu}^{(c)}$ labels all the hopping matrix
elements within a cluster and the above equation defines $V_{\mu\nu}^{mn}$.

We pause to introduce the notation that will be used throughout for quantum
cluster methods. We follow the review article Ref.~\onlinecite{Maier:2005}. In
reciprocal space, any wave vector $\mathbf{k}$ in the Brillouin zone may be
written as $\mathbf{k=}\tilde{\mathbf{k}}+\mathbf{K}$ where both
$\mathbf{k}$ and $\tilde{\mathbf{k}}$ are continuous in the infinite size
limit, except that $\tilde{\mathbf{k}}$ is defined only in the reduced
Brillouin zone that corresponds to the superlattice. On the other hand,
$\mathbf{K}$ is discrete and denotes reciprocal lattice vectors of the
superlattice. By analogy, any position $\mathbf{r}$ in position space can be
written as $\tilde{\mathbf{r}}+\mathbf{R}$ where $\mathbf{R}$ is for
positions within clusters while $\tilde{\mathbf{r}}$ labels the origins of
the clusters, an infinite number of them. Hence, Fourier's theorem allows one
to define functions of $\mathbf{k,}$ $\tilde{\mathbf{k}}$ or $\mathbf{K}$
that contain the same information as functions of, respectively, $\mathbf{r,}$
$\tilde{\mathbf{r}}$ or $\mathbf{R.}$ Also, we have $\mathbf{K\cdot}$
$\tilde{\mathbf{r}}=2\pi n$ where $n$ is an integer. Sites within a
cluster are labelled by greek letters so that the position of site $\mu$
within a cluster is $\mathbf{R}_{\mu}$, while clusters are labelled by Latin
letters so that the origin of cluster $m$ is at $\tilde{\mathbf{r}}_{m}$.

Returning to CPT, the Green function for the whole system is given by
\begin{equation}
\left[\hat{G}^{-1}(\tilde{\mathbf{k}},z)\right]  _{\mu\nu}=\left[
\hat{G}^{(c)-1}(z)-\hat{V}(\tilde{\mathbf{k}})\right]  _{\mu\nu}
\label{Gsuperlattice}%
\end{equation}
where hats denote matrices in cluster site indices and $z$ is the complex
frequency. At this level of approximation, the CPT Green function has the same
structure as in the Hubbard I approximation except that it pertains to a
cluster instead of a single site. Since $\hat{G}^{(c)-1}(z)=z+\mu
-\hat{t}^{(c)}-\hat{\Sigma}^{(c)}$ and $\hat{G}^{(0)-1}%
(\tilde{\mathbf{k}},z)=z+\mu-\hat{t}^{(c)}-$ $\hat{V}%
(\tilde{\mathbf{k}}),$ the Green function (\ref{Gsuperlattice}) may
also be written as
\begin{equation}
\hat{G}^{-1}(\tilde{\mathbf{k}}\mathbf{,}z)=\hat{G}^{(0)-1}%
(\tilde{\mathbf{k}}\mathbf{,}z)-\hat{\Sigma}^{(c)}(z).
\end{equation}
This form allows a different physical interpretation of the approach. In the
above expression, the self-energy of the lattice is approximated by the
self-energy of the cluster. The latter in real space spans only the size of
the cluster.

We still need an expression to extend the above result to the lattice in a
translationally invariant way.
This is done by defining the following residual Fourier transform:
\begin{equation}
G_{\rm CPT}(\mathbf{k},z)=\frac{1}{N_{c}}\sum_{\mu,\nu}^{N_{c}}e^{i\mathbf{k}\cdot(\mathbf{R}_{\mu}-\mathbf{R}_{\nu})}G_{\mu\nu}(\tilde{\mathbf{k}},z).
\label{GCPT}%
\end{equation}
Notice that $G_{\mu\nu}(\tilde{\mathbf{k}},z)$ may be replaced by
$G_{\mu\nu}(\mathbf{k},z)$ in the above equation since $\hat{V}(\tilde{\mathbf{k}}+\mathbf{K}) = \hat{V}(\tilde{\mathbf{k}})$.

\subsubsection{Self-energy functional approach}

The self-energy functional approach, devised by Potthoff\cite{Potthoff:2003}
allows one to consider various cluster schemes from a unified point of view.
It begins with $\Omega_{\mathbf{t}}[G],$ a functional of the Green function%
\begin{equation}
\Omega_{\mathbf{t}}[G]=\Phi[G]-\mathrm{Tr}((G_{0\mathbf{t}}^{-1}%
-G^{-1})G)+\mathrm{Tr}\ln(-G). \label{GrandPotential}%
\end{equation}
The Luttinger Ward functional $\Phi[G]$ entering this equation is the
sum of connected vacuum skeleton diagrams. A diagram-free definition of this
functional is also given in Ref.~\onlinecite{Potthoff:2004}. For our purposes,
what is important is that (1) The functional derivative of $\Phi[G]$ is
the self-energy%
\begin{equation}
\frac{\delta\Phi[G]}{\delta G}=\Sigma\label{SelfLuttinger}%
\end{equation}
and (2) it is a universal functional of $G$ in the following sense: whatever
the form of the one-body Hamiltonian, it depends only on the interaction and,
functionnally, it has the same dependence
on $G$. The dependence of the functional $\Omega_{\mathbf{t}%
}[G]$ on the one-body part of the Hamiltonian is denoted by the subscript
$\mathbf{t}$ and it comes only through $G_{0\mathbf{t}}^{-1}$ appearing on the
right-hand side of Eq.~(\ref{GrandPotential}).

The functional $\Omega_{\mathbf{t}}[G]$ has the important property that it is
stationary when $G$ takes the value prescribed by Dyson's equation. Indeed,
given the last two equations, the Euler equation takes the form%
\begin{equation}
\frac{\delta\Omega_{\mathbf{t}}[G]}{\delta G}=\Sigma-G_{0\mathbf{t}}%
^{-1}+G^{-1}=0.
\end{equation}
This is a dynamic variational principle since it involves the frequency
appearing in the Green function, in other words excited states are involved in
the variation. At this stationary point, and only there, $\Omega_{\mathbf{t}%
}[G]$ is equal to the grand potential. Contrary to Ritz's variational
principle, this last equation does not tell us whether $\Omega_{\mathbf{t}%
}[G]$ is a minimum or a maximum or a saddle point there.

There are various ways to use the stationarity property that we described
above. The most common one, is to approximate $\Phi[G]$ by a finite set
of diagrams. This is how one obtains the Hartree-Fock, the FLEX
approximation\cite{Bickers:1989} or other so-called thermodynamically
consistent theories. This is what Potthoff calls a type II
approximation strategy.\cite{Potthoff:2005} A type I approximation simplifies the Euler equation
itself. In a type III approximation, one uses the exact form of $\Phi[
G]$ but only on a limited domain of trial Green functions.

Following Potthoff, we adopt the type III approximation on a functional
of the self-energy instead of on a functional of the Green function. Suppose
we can locally invert Eq.~(\ref{SelfLuttinger}) for the self-energy  to write $G$ as a functional of $\Sigma.$ We can use this result to write,
\begin{equation}
\Omega_{\mathbf{t}}[\Sigma]=F[\Sigma]-\mathrm{Tr}\ln(-G_{0\mathbf{t}}%
^{-1}+\Sigma).
\end{equation}
where we defined
\begin{equation}
F[\Sigma]=\Phi[G]-\mathrm{Tr}(\Sigma G).
\end{equation}
and where it is implicit that $G=G[\Sigma]$ is now a functional of $\Sigma$.
$F[\Sigma],$ along with the expression (\ref{SelfLuttinger}) for the derivative of the Luttinger-Ward functional, define the Legendre
transform of the Luttinger-Ward functional. It is easy to verify that%
\begin{equation}
\frac{\delta F[\Sigma]}{\delta\Sigma}=\frac{\delta\Phi[G]}{\delta
G}\frac{\delta G[\Sigma]}{\delta\Sigma}-\Sigma\frac{\delta G[\Sigma]}%
{\delta\Sigma}-G=-G
\end{equation}
hence, $\Omega_{\mathbf{t}}[\Sigma]$ is stationary with respect to $\Sigma$
when Dyson's equation is satisfied%
\begin{equation}
\frac{\delta\Omega_{\mathbf{t}}[\Sigma]}{\delta\Sigma}=-G+(G_{0\mathbf{t}}%
^{-1}-\Sigma)^{-1}=0.
\end{equation}

\begin{figure}[ptb]
\centerline{\includegraphics[width=8.5cm]{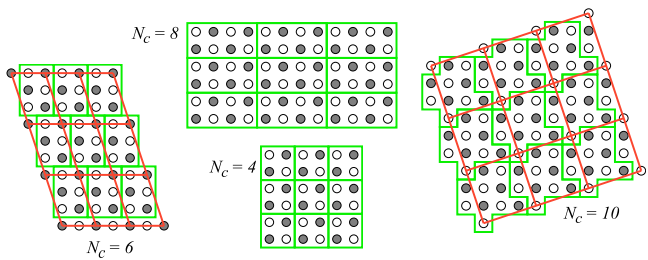}} \caption{Various tilings used
in quantum cluster approaches. In these examples the grey and white sites are
inequivalent since an antiferromagnetic order is possible. }%
\label{fig_42}%
\end{figure}

To perform a type III approximation on $F[\Sigma]$, we take advantage that it
is universal, i.e., that it depends only on the interaction part of the
Hamiltonian and not on the one-body part. This follows from the universal
character of its Legendre transform $\Phi[G]$. We thus evaluate
$F[\Sigma]$ exactly for a Hamiltonian $H^{\prime}$ that shares the same
interaction part as the Hubbard Hamiltonian, but that is exactly solvable.
This Hamiltonian $H^{\prime}$ is taken as a cluster decomposition of the
original problem, i.e., we tile the infinite lattice into identical,
disconnected clusters that can be solved exactly. Examples of such tilings are
given in Fig.~\ref{fig_42}. Denoting the corresponding quantities with a
prime, we obtain,
\begin{equation}
\Omega_{\mathbf{t}^{\prime}}[\Sigma^{\prime}]=F[\Sigma^{\prime}]-\mathrm{Tr}%
\ln(-G_{0\mathbf{t}^{\prime}}^{-1}+\Sigma^{\prime}).
\end{equation}
from which we can extract $F[\Sigma^{\prime}]$. It follows that
\begin{equation}
\Omega_{\mathbf{t}}[\Sigma^{\prime}]=\Omega_{\mathbf{t}^{\prime}}%
[\Sigma^{\prime}]+\mathrm{Tr}\ln(-G_{0\mathbf{t}^{\prime}}^{-1}+\Sigma
^{\prime})-\mathrm{Tr}\ln(-G_{0\mathbf{t}}^{-1}+\Sigma^{\prime}).
\label{sef_eq}%
\end{equation}
The type III approximation comes from the fact that the self-energy
$\Sigma^{\prime}$ is restricted to the exact self-energy of the cluster
problem $H^{\prime}$, so that variational parameters appear in the definition
of the one-body part of $H^{\prime}$.

In practice, we look for values of the cluster one-body parameters
$\mathbf{t}^{\prime}$ such that $\delta\Omega_{\mathbf{t}}[\Sigma^{\prime
}]/\delta\mathbf{t}^{\prime}=0$. It is useful for what follows to write the
latter equation formally, although we do not use it in actual calculations.
Given that $\Omega_{\mathbf{t}^{\prime}}[\Sigma^{\prime}]$ is the actual grand
potential evaluated for the cluster, $\partial\Omega_{\mathbf{t}^{\prime}%
}[\Sigma^{\prime}]/\partial\mathbf{t}^{\prime}$ is canceled by the explicit
$\mathbf{t}^{\prime}$ dependence of $\mathrm{Tr}\ln(-G_{0\mathbf{t}^{\prime}%
}^{-1}+\Sigma^{\prime})$ and we are left with
\begin{align}
0  &  =\frac{\delta\Omega_{\mathbf{t}}[\Sigma^{\prime}]}{\delta\Sigma^{\prime
}}\frac{\delta\Sigma^{\prime}}{\delta\mathbf{t}^{\prime}}\nonumber\\
&  =-\mathrm{Tr}\left[\left(  \frac{1}{G_{0\mathbf{t}^{\prime}}^{-1}%
-\Sigma^{\prime}}-\frac{1}{G_{0\mathbf{t}}^{-1}-\Sigma^{\prime}}\right)
\frac{\delta\Sigma^{\prime}}{\delta\mathbf{t}^{\prime}}\right]  .
\end{align}
Given that the clusters corresponding to $\mathbf{t}^{\prime}$ are
disconnected and that translation symmetry holds on the superlattice of
clusters, each of which contains $N_c$ sites,
the last equation may be written
\begin{align}
&  \sum_{\omega_{n}}\sum_{\mu\nu}\bigg[\frac{N}{N_{c}}\left(  \frac
{1}{G_{0\mathbf{t}^{\prime}}^{-1}-\Sigma^{\prime}(i\omega_{n})}\right)
_{\mu\nu}\nonumber\\
&  ~~ -\sum_{\tilde{\mathbf{k}}}\left(  \frac{1}{G_{0\mathbf{t}}%
^{-1}(\tilde{\mathbf{k}})-\Sigma^{\prime}(i\omega_{n})}\right)  _{\mu\nu
}\bigg]\frac{\delta\Sigma_{\nu\mu}^{\prime}(i\omega_{n})}{\delta
\mathbf{t}^{\prime}}=0. \label{EulerVCA}%
\end{align}

\subsubsection{Variational cluster perturbation theory, or variational cluster
approximation}

In Variational Cluster Perturbation Theory (VCPT), more aptly named the
Variational Cluster Approach (VCA), solutions to the Euler equations
(\ref{EulerVCA}) are found by looking for numerical minima (or more generally,
saddle-points) of the functional. Typically, the VCA
cluster Hamiltonian $H^{\prime}$ will have the same form as $H$ except that
there is no hopping between clusters and that long-range order is allowed by
adding some Weiss fields, for instance like in Eq.~(\ref{weiss_eq}) below. The
hopping terms and chemical potential within $H^{\prime}$ may also be treated
like additional variational parameters. In contrast with Mean-Field theory,
these Weiss fields are not mean fields, in the sense that they do not coincide
with the corresponding order parameters. The interaction part of $H$ (or $H$')
is not factorized in any way and short-range correlations are treated exactly.
In fact, the Hamiltonian $H$ is not altered in any way; the Weiss fields are
introduced to let the variational principle act on a space of self-energies
that includes the possibility of specific long-range orders, without imposing
those orders. Indeed, the more naturally an order arises in the system, the
smaller the Weiss field needs to be, and one observes that the strength of the
Weiss field at the stationary point of the self-energy functional generally
decreases with increasing cluster size, as it should since in the thermodynamic limit no
Weiss field should be necessary to establish order.

\subsubsection{Cellular dynamical mean-field theory}

The Cellular dynamical mean-field theory (CDMFT) is obtained by including in
the cluster Hamiltonian $H^{\prime}$ a bath of uncorrelated electrons that
somehow must mimic the effect on the cluster of the rest of the lattice.
Explicitly, $H^{\prime}$ takes the form
\begin{align}
H^{\prime} &  =-\sum_{\mu,\nu,\sigma}t_{\mu\nu}^{\prime}c_{\mu\sigma}%
^{\dagger}c_{\nu\sigma}+U\sum_{\mu}n_{\mu\uparrow}n_{\mu\downarrow}\nonumber\\
&  ~~+\sum_{\mu,\alpha,\sigma}V_{\mu\alpha}(c_{\mu\sigma}^{\dagger}%
a_{\alpha\sigma}+\mathrm{H.c.})+\sum_{\alpha}\epsilon_{\alpha}a_{\alpha\sigma
}^{\dagger}a_{\alpha\sigma}%
\end{align}
where $a_{\alpha\sigma}$ annihilates an electron of spin $\sigma$ on a bath
orbital labelled $\alpha$. The bath is characterized by the energy of each
orbital ($\epsilon_{\alpha}$) and the bath-cluster hybridization matrix
$V_{\mu\alpha}$. This representation of the environment through an Anderson
impurity model was introduced in Ref.~\onlinecite{Caffarel:1994} in the
context of DMFT (i.e., a single site). The effect of the bath on the electron
Green function is encapsulated in the so-called hybridization function
\begin{equation}
\Gamma_{\mu\nu}(\omega)=\sum_{\alpha}{\frac{V_{\mu\alpha}V_{\nu\alpha}^{\ast}%
}{\omega-\epsilon_{\alpha}}}%
\end{equation}
which enters the Green function as
\begin{equation}
[G^{\prime-1}]_{\mu\nu}=\omega+\mu-t_{\mu\nu}^{\prime}-\Gamma_{\mu\nu
}(\omega)-\Sigma_{\mu\nu}(\omega).
\end{equation}

Moreover, the CDMFT does not look for a strict solution of the Euler equation
(\ref{EulerVCA}), but tries instead to set each of the terms between brackets to zero separately. Since the Euler
equation (\ref{EulerVCA}) can be seen as a scalar product, CDMFT requires
that the modulus of one of the vectors vanish to make the scalar product
vanish. From a heuristic point of view, it is as if each component of the
Green function in the cluster were equal to the corresponding component
deduced from the lattice Green function. This clearly reduces to single site
DMFT when there is only one lattice site.

When the bath is discretized, i.e., is made of a finite number of bath
\textquotedblleft orbitals\textquotedblright, the left-hand side of
Eq.~(\ref{EulerVCA}) cannot vanish separately for each frequency, since the
number of degrees of freedom in the bath is insufficient. Instead, one adopts
the following self-consistent scheme: (1) one starts with a guess value of the
bath parameters $(V_{\mu\alpha},\epsilon_{\alpha})$ and solves the cluster
Hamiltonian $H^{\prime}$ numerically. (2) One then calculates the combination
\begin{equation}
\hat{\mathcal{G}}_{0}^{-1}=\left[\sum_{\tilde{\mathbf{k}}}\frac
{1}{\hat{G}_{0\mathbf{t}}^{-1}(\tilde{\mathbf{k}})-\hat{\Sigma}^{\prime}(i\omega_{n}%
)}\right]  ^{-1}+\hat\Sigma^{\prime}(i\omega_{n})
\end{equation}
and (3) minimizes the following canonically invariant distance function:
\begin{equation}
d=\sum_{n,\mu,\nu}\left\vert\left( i\omega_{n}+\mu-\hat{t}^\prime-\hat\Gamma
(i\omega_{n})-\hat{\mathcal{G}}_{0}^{-1}\right)_{\mu\nu}\right\vert ^{2}\label{dist_func}%
\end{equation}
over the set of bath parameters (changing the bath parameters at this step
does not require a new solution of the Hamiltonian $H^{\prime}$, but merely a
recalculation of the hybridization function $\hat{\Gamma}$). The bath
parameters obtained from this minimization are then put back into step (1) and
the procedure is iterated until convergence.

In practice, the distance function (\ref{dist_func}) can take various forms,
for instance by adding a frequency-dependent weight in order to emphasize
low-frequency properties\cite{Kancharla:2005, Bolech:2003, Stanescu:2005} or
by using a sharp frequency cutoff.\cite{Kyung:2005} These weighting factors
can be considered as rough approximations for the missing factor $\delta
\Sigma_{\nu\mu}^{\prime}(i\omega_{n})/\delta\mathbf{t}^{\prime}$ in the Euler
equation (\ref{EulerVCA}). The frequencies are summed over on a discrete,
regular grid along the imaginary axis, defined by some fictitious inverse
temperature $\beta$, typically of the order of 20 or 40 (in units of $t^{-1}%
$). Even when the total number of cluster plus bath sites in CDMFT equals the
number of sites in a VCA calculation, CDMFT is much faster than
the VCA since the minimization of a grand potential functional requires many exact diagonalizations of the cluster Hamiltonian $H'$.

The final lattice Green function from which one computes observable quantities
may be obtained by periodizing the self-energy, as in
Ref.~\onlinecite{Kotliar:2001} or in the CPT manner described above in
Eq.~(\ref{GCPT}). We prefer the last approach because it corresponds to the
Green function needed to obtain the density from $\partial\Omega/\partial
\mu=-\mathrm{Tr}(G)$ and also because periodization of the self-energy gives
additional unphysical states in the Mott gap\cite{Senechal:2003} (see also Ref.~\onlinecite{Stanescu:2004}).

\subsubsection{The Dynamical cluster approximation}

The DCA\cite{Hettler:1998} cannot be formulated within the self-energy
functional approach.\footnote{Th. Maier, M. Potthoff and D. S\'{e}n\'{e}chal,
unpublished.}
It is based on the idea of discretizing irreducible quantities,
such as the energy, in reciprocal space. It is believed to converge faster for
$\mathbf{q=0}$ quantities whereas CDMFT converges exponentially fast for local
quantities.\cite{Biroli:2002, Aryanpour:2005, Biroli:2005}

\subsubsection{Benchmarks for quantum cluster approaches}

Since DMFT becomes exact in infinite dimension, the most difficult challenge
for cluster extensions of this approach is in one dimension. In addition,
exact results to compare with exist only in one dimension so it is mostly in
$d=1$ that cluster methods have been checked. In $d=2$ there have also been a
few comparisons with QMC as we shall discuss.

\begin{figure}[t]
\centerline {\includegraphics[width=7cm]{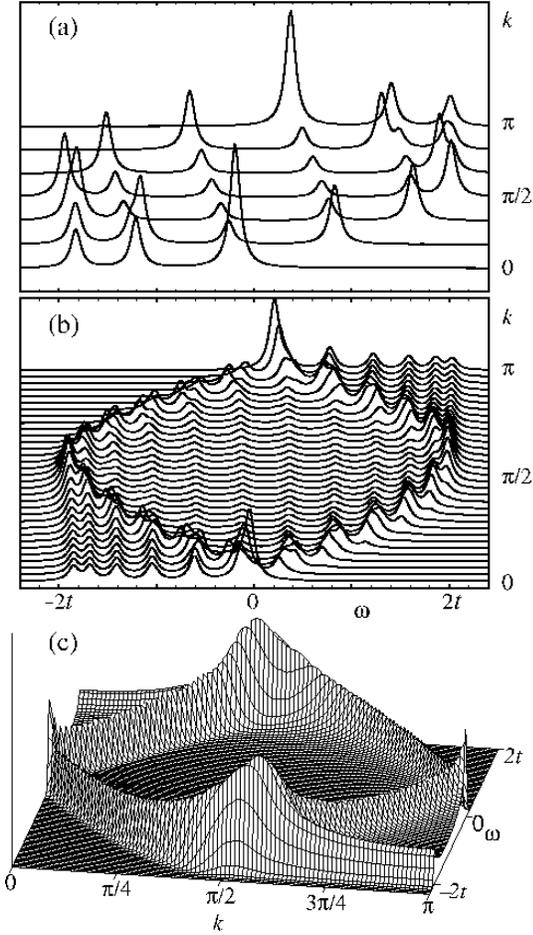}}\caption{The spectral function
of the $U\to\infty$ limit of the one- dimensional Hubbard model, as calculated
from (a) an exact diagonalization of the Hubbard model with $U/t=100$ on a
periodic 12- site cluster; (b) the same, but with CPT, on a 12-site cluster
with open boundary conditions; (c) the exact solution, taken from
Ref.~\onlinecite{Favand:1997}; beware: the axes are oriented differently. In
(a) and (b) a finite width $\eta$ has been given to peaks that would otherwise
be Dirac $\delta$-functions.}%
\label{fig_53a}%
\end{figure}

\begin{figure}[ptb]
\centerline{\includegraphics[width=6.5cm]{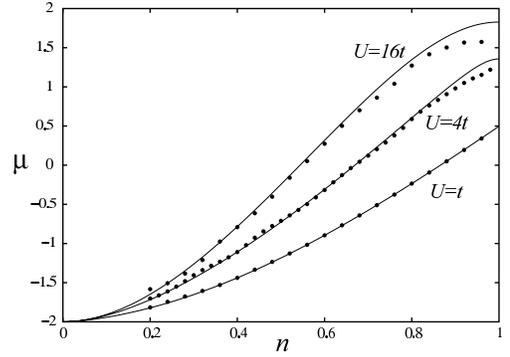}} \caption{Chemical potential
as a function of density in the one-dimensional Hubbard model, as calculated
by CPT (from Ref.~\onlinecite{Senechal:2002}). The exact, Bethe-Ansatz result
is shown as a solid line.}%
\label{fig_53b}%
\end{figure}

\begin{figure}[ptb]
\centerline{\includegraphics[width=6.5cm]{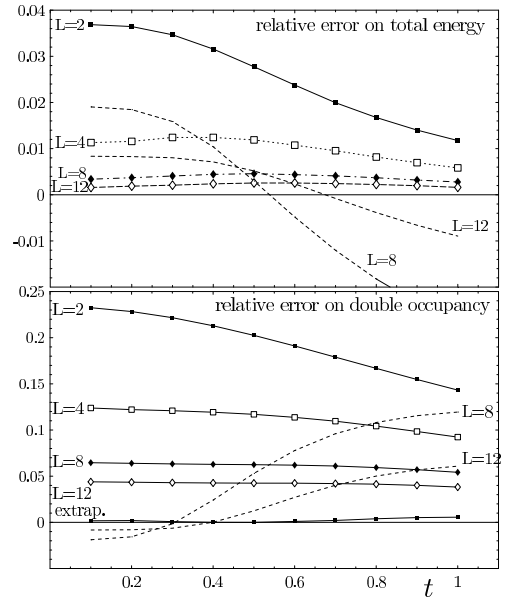}} \caption{Top: Comparison
(expressed in relative difference) between the ground-state energy density of
the half-filled, one-dimensional Hubbard model calculated from the exact,
Bethe-Ansatz result. The results are displayed as a function of the hopping
$t$, for $U=2t$ and various cluster sizes $L$ (connected symbols). For
comparison, the exact diagonalization values of finite clusters with periodic
boundary conditions are also shown (dashed lines) for $L=8$ and $L=12$.
Bottom: Same for the double occupancy. An extrapolation of the results to
infinite cluster size ($L\to\infty$) using a quadratic fit in terms of $1/L$
is also shown, and is accurate to within 0.5\%. Taken from
Ref.~\onlinecite{Senechal:2002}.}%
\label{fig_54}%
\end{figure}

CPT has been checked\cite{Senechal:2003} for example by comparing
with exact results\cite{Favand:1997}
for the spectral function at $U\rightarrow\infty$ in $d=1$
as shown in Fig.~\ref{fig_53a}. Fig.~\ref{fig_53b} shows the chemical potential
as a function of density for various values of $U$. Fig.~\ref{fig_54} shows
the convergence rates for the total energy and for the double occupancy in the
$d=1$ half-filled model. Clearly, there is a dramatic improvement compared
with exact diagonalizations.

\begin{figure}[ptb]
\centerline{\includegraphics[width=5.5cm]{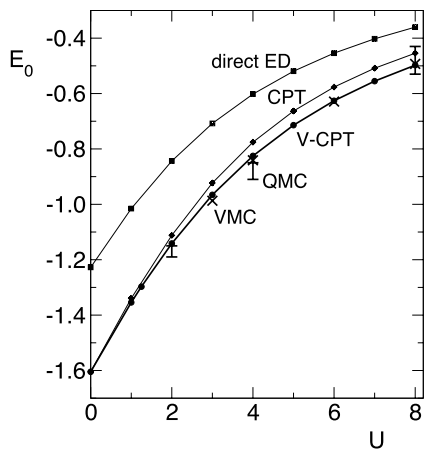}} \caption{Ground state energy
of the half-filled, two-dimensional Hubbard model ($t=1$) as a function of
$U$, as obtained from various methods: Exact diagonalization (ED), CPT and
VCPT on a 10-site cluster, quantum Monte Carlo (QMC) and variational Monte
Carlo (VMC). Taken from Ref.~\onlinecite{Dahnken:2004}.}%
\label{fig_55b}%
\end{figure}

The main weakness of CPT is that it cannot take into account tendency towards
long-range order. This is remedied by VCPT, as shown in Fig.~\ref{fig_55b}
where CPT, VCPT are both compared with QMC as a benchmark. Despite this
agreement, we should stress that long wave length fluctuations are clearly
absent from cluster approaches. Hence, the antiferromagnetic order parameter
at half-filling, for example, does not contain the effect of zero-point long
wave length transverse spin fluctuations. This is discussed for example in the
context of Fig.~9 of Ref.~\onlinecite{Dahnken:2004}.

\begin{figure}[ptb]
\centerline{\includegraphics[width=6.5cm]{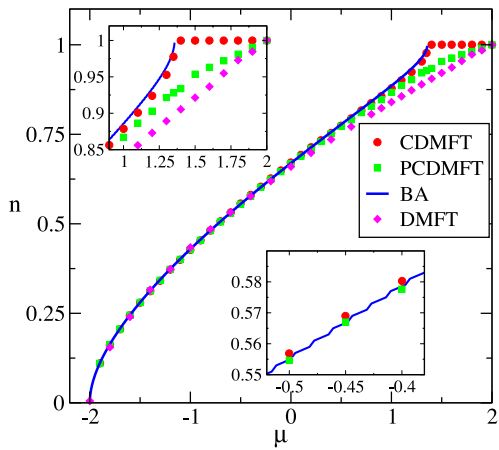}} \caption{CDMFT calculation on
a $2 \times2$ cluster with $8$ bath sites of the density as a function of the
chemical potential in the one-dimensional Hubbard model for $U=4t$, as
compared with the exact solution, DMFT and other approximation schemes. Taken
from Ref.~\onlinecite{Capone:2004}.}%
\label{fig_56}%
\end{figure}

CDMFT corrects the difficulties of CPT near half-filling by reproducing the
infinite compressibility predicted by the Bethe ansatz in one dimension as
shown in Fig.~\ref{fig_56}.\cite{Capone:2004} Detailed comparisons between the
local and near-neighbor Green functions\cite{Capone:2004, Bolech:2003} have
been performed. One should note that these results, obtained from exact
diagonalization, also need the definition of a distance function (See
Eq.~(\ref{dist_func}) above) that helps find the best bath parametrization to
satisfy the self-consistency condition. This measure forces one to define
calculational parameters such as a frequency cutoff and an fictitious
temperature defining the Matsubara frequencies to sum over. The final results
are not completely insensitive to the choice of fictitious temperature or
frequency-weighing scheme but are usually considered reliable and consistent
with each other when $\beta$ lies between $20$ and $40$. The cutoff procedures
have been discussed in Ref.~\onlinecite{Kyung:2005}.

The relative merits of DCA and CDMFT have been discussed for example in Refs~
\onlinecite{Biroli:2002, Aryanpour:2005, Biroli:2005, Maier:2002a, Pozgajcic:2004}.
Briefly speaking, convergence seems faster in DCA for long wave length
quantities but CDMFT is faster (exponentially) for local quantities.

\section{Results and concordance between different methods\label{Results}}

In this section, we outline the main results we obtained concerning the
pseudogap and d-wave superconductivity in the two-dimensional Hubbard model.
Quantum cluster approaches are better at strong coupling while TPSC is best at
weak coupling. Nevertheless, all these methods are non-perturbative, the
intermediate coupling regime presenting the physically most interesting case.
But it is also the regime where we have the least control over the
approximations. As we will show, it is quite satisfying that, at
intermediate coupling, weak-coupling and strong-coupling approaches give
results that are nearly in quantitative agreement with each other. This gives
us great confidence into the validity of the results. As an example of
concordance, consider the fact that to obtain spectral weight near $(\pi
/2,\pi/2)$ at optimal doping in the electron-doped systems, $U$ has to be
roughly $6t$. For larger $U,$ ($U=8t$ in CPT) that weight, present in the experiments,
disappears. Smaller values of $U$ ($U=4t$ in CPT)
do not lead to a pseudogap. Other examples
of concordance include the value of the superconducting transition
temperature $T_{c}$ obtained with DCA and with TPSC as well as the
temperature dependence of souble occupancy obtained with the same
two methods.

\subsection{Weak and strong-coupling pseudogap\label{WeakAndStrongCoupling}}

To understand the pseudogap it is most interesting to consider both hole and
electron-doped cuprates at once. This means that we have to include
particle-hole symmetry breaking hoppings, $t^{\prime}$ and $t^{\prime\prime}$.
We will see in the present section that it is possible to obtain a pseudogap
at strong coupling without a large correlation length in the particle-hole or
in the particle-particle channels. By contrast, at weak coupling one does need
a long-correlation length and low dimension. So there appears to be
theoretically two different mechanisms for the pseudogap.

\begin{figure}[ptb]
\centerline{\includegraphics[width=6.5cm]{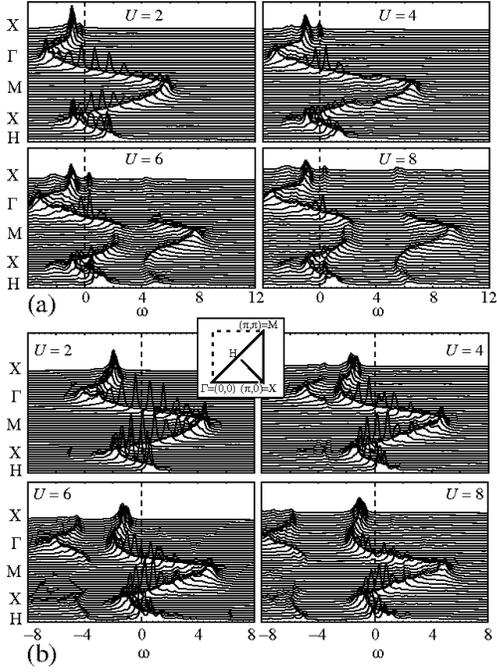}} \caption{ Single particle
spectral weight, as a function of energy $\omega$ in units of $t$, for wave
vectors along the high-symmetry directions shown in the inset. (a) CPT
calculations on a $3 \times4$ cluster with ten electrons (17\% hole doped).
(b) the same as (a), with 14 electrons (17\% electron doped). In all cases
$t^{\prime}=-0.3t$ and $t^{\prime\prime}=0.2t$. A Lorentzian broadening
$\eta=0.12t$ is used to reveal the otherwise delta peaks. From
Ref.~\onlinecite{Senechal:2004}.}%
\label{fig_62}%
\end{figure}

\begin{figure}[ptb]
\centerline{\includegraphics[width=6.5cm]{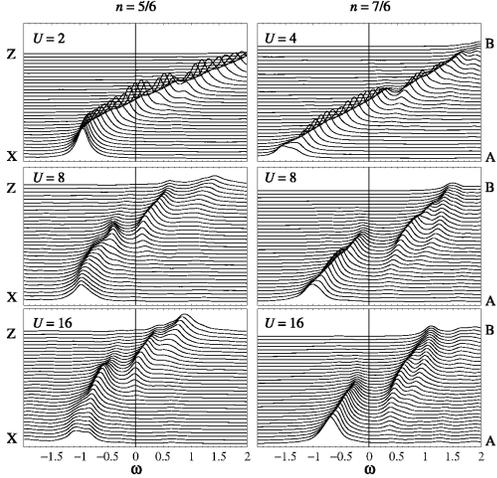}} \caption{Onset of the
pseudogap as a function of $U$ corresponding to Fig.~\ref{fig_62}, taken from
Ref.~\onlinecite{Senechal:2004}. Hole-doped case on the left, electron-doped
case on the right panel}%
\label{fig_63}%
\end{figure}

\subsubsection{CPT}

The top panel in Fig.~\ref{fig_62} presents the single-particle spectral
weight, $A(\mathbf{k},\omega)$ or imaginary part of the single-particle Green
function, for the model with $t^{\prime}=-0.3t,$ $t^{\prime\prime}=0.2t$ in
the hole-doped case, for about $17\%$ doping.\cite{Senechal:2004} Each curve
is for a different wave vector (on a trajectory shown in the inset) and is
plotted as a function of frequency in units of $t$. This kind of plot is known
as Energy Dispersion Curves (EDC). It is important to point out that the
theoretical results are obtained by broadening a set of delta function, so
that the energy resolution is $\eta=0.12t$ corresponding roughly to the
experimental resolution we will compare with in the next section. At small
$U=2t$ on the top panel of Fig.~\ref{fig_62}, one recovers a Fermi liquid. At
large $U$, say $U=8t$, the Mott gap at positive energy is a prominent feature.
The pseudogap is a different feature located around the Fermi energy. To see
it better, we present on the left-hand panel of Fig.~\ref{fig_63} a blow-up in
the vicinity of the Fermi surface crossing occurring near $(\pi,0)$.
Clearly, there is a minimum in $A(\mathbf{k},\omega)$ at the
Fermi-surface crossing when $U$ is large enough instead of a maximum like in
Fermi liquid theory.

\begin{figure}[ptb]
\centerline{\includegraphics[width=6.5cm]{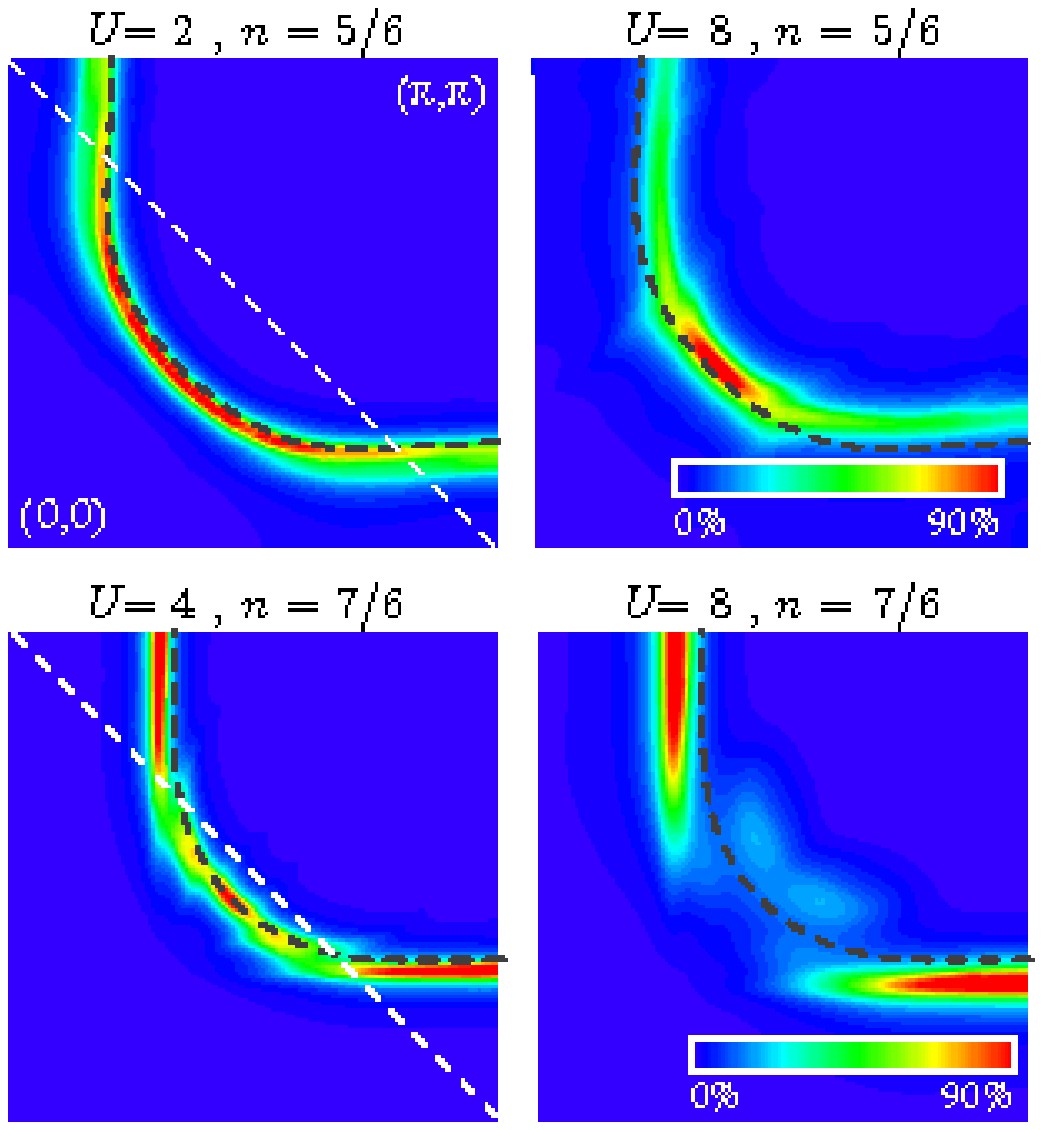}} \caption{MDC from CPT in the
$t$-$t^{\prime}=-0.3t$, $t^{\prime\prime}=0.2t$ Hubbard model, taken from
Ref.~\onlinecite{Senechal:2004}. }%
\label{fig_64}%
\end{figure}

It is also possible to plot $A(\mathbf{k},\omega)$ at fixed frequency for
various momenta. They are so-called Momentum Dispersion Curves (MDC). In
Fig.~\ref{fig_64} we take the Fermi energy $\omega=0,$ and we plot the
magnitude of the single-particle spectral weight in the first quadrant of the
Brillouin zone using red for high-intensity and blue for low intensity. The
figure shows that, in the hole-doped case (top panel), weight near $(\pi
/2,\pi/2)$ survives while it tends to disappear near $(\pi,0)$ and $(0,\pi)$.
That pseudogap phenomenon is due not only to large $U$ but also to the fact
that the line that can be drawn between the points $(\pi,0)$ and $(0,\pi)$
crosses the Fermi surface. When there is no such crossing, one recovers a
Fermi surface (not shown here). The $(\pi,0)$ to $(0,\pi)$ line has a double
role. It is the antiferromagnetic zone boundary, as well as the line that
indicates where umklapp processes become possible, i.e., the line where we can
scatter a pair of particles on one side of the Fermi surface to the other side
with loss or gain of a reciprocal lattice vector. Large scattering rates
explain the disappearance of the Fermi surface.\cite{Senechal:2004} We also
note that the size of the pseudogap in CPT, defined as the distance between
the two peaks, does not scale like $J=4t^{2}/U$ at large coupling. It seems to
be very weakly $U$ dependent, its size being related to $t$ instead. This
result is corroborated by CDMFT.\cite{Kyung:2005}

\begin{figure}[ptb]
\centerline{\includegraphics[width=8.5cm]{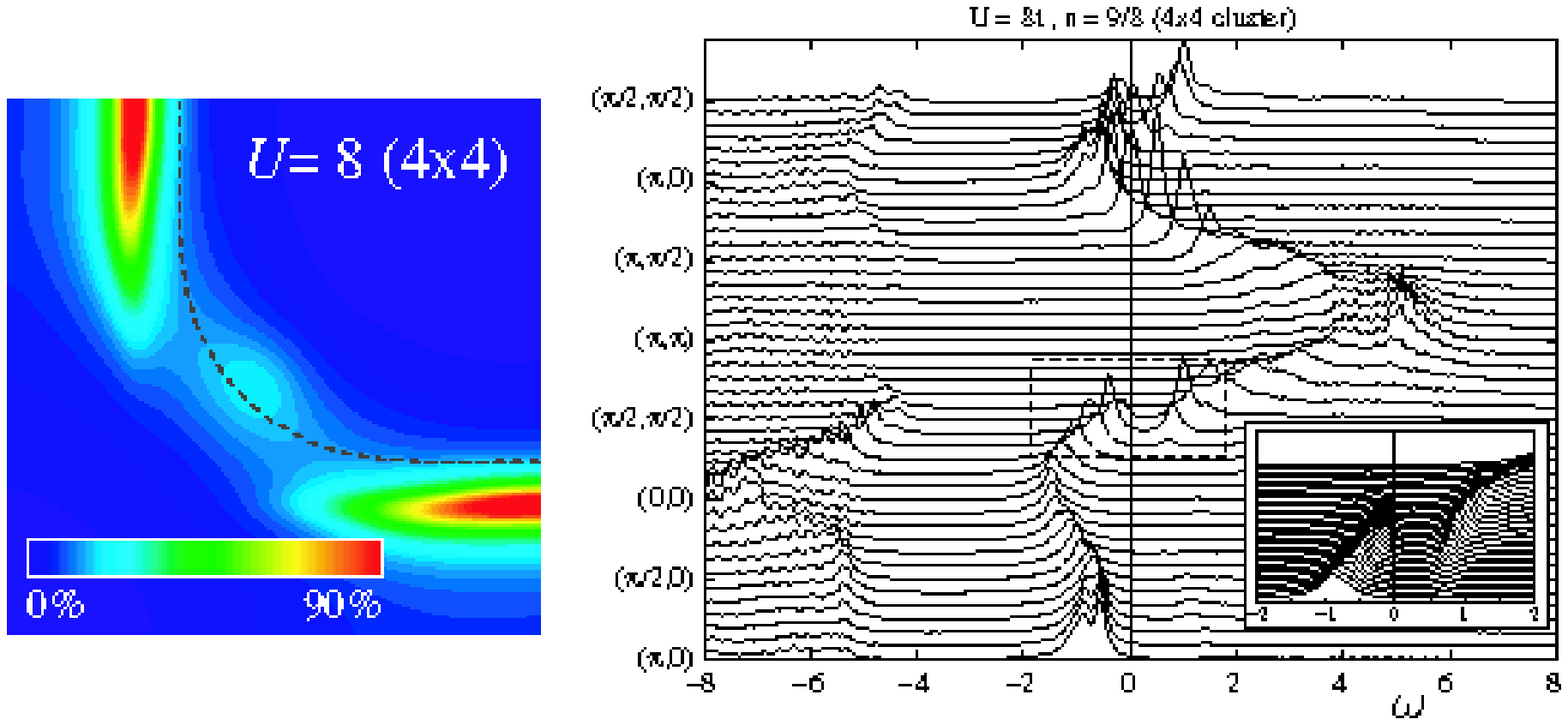}} \caption{Right: The corresponding EDC in the $t$-$t^{\prime}$-$t^{\prime\prime}$ Hubbard model, calculated on a 16-site
cluster in CPT, at $n=9/8$. Inset: the pseudogap. Left: The corresponding Fermi energy
momentum distribution curve.}%
\label{fig_71}%
\end{figure}

The EDC for the electron-doped case is shown on the bottom panels of
Fig.~\ref{fig_62} near optimal doping again. This time, the Mott gap appears
below the Fermi surface so that the lower Hubbard band becomes accessible to
experiment. The EDC in Fig.~\ref{fig_71} shows very well both the Mott gap and
the pseudogap. Details of that pseudogap can be seen both in the inset of Fig.
\ref{fig_71} or on the right-hand panel of Fig. \ref{fig_63}. While in the
hole-doped case the MDC appeared to evolve continuously as we increase $U$
(top panel of Fig.~\ref{fig_64}), in the electron-doped case (bottom panel)
the weight initially present near $(\pi/2,\pi/2)$ at $U=4t$ disappears by the
time we reach $U=8t$.

\begin{figure}[ptb]
\centerline{\includegraphics[width=6.5cm]{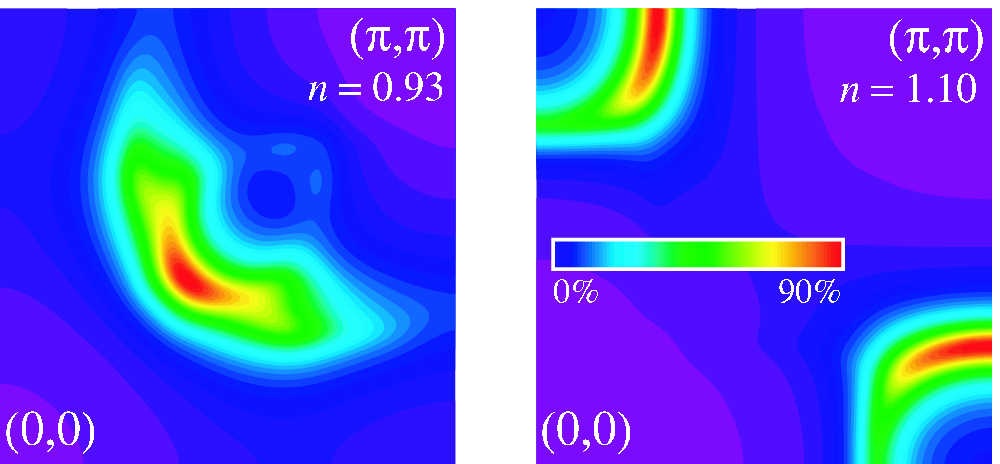}} \caption{EDC in the
$t$-$t^{\prime}$-$t^{\prime\prime}$ Hubbard model, with $t'=-0.3t$ and $t''=0.2t$, calculated on a
8-site cluster for $U=8t$ in VCPT. d-wave superconductivity is present in the
holde-doped case (left) and both antiferromagnetism and d-wave
superconductivity in the electron-doped case. The resolution is not large
enough in the latter case to see the superconducting gap. The Lorentzian
broadening is $0.2t$. From Ref.~\onlinecite{Senechal:2005}. }%
\label{fig_110}%
\end{figure}

In Fig.~\ref{fig_110} we show, with the same resolution as CPT, the MDC for
VCPT.\cite{Senechal:2005} In this case the effect of long-range order is
included and visible but, at this resolution, the results are not
too different from those
obtained from CPT in Fig.~\ref{fig_64}.

\begin{figure}[ptb]
\centerline{\includegraphics[width=7.5cm]{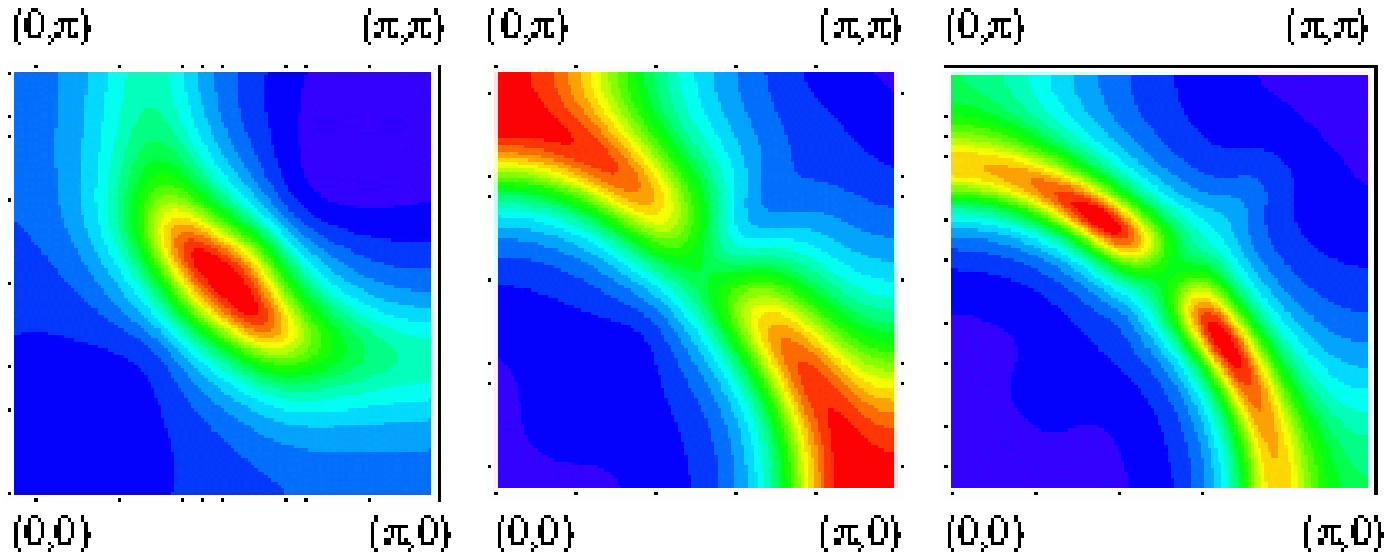}} \caption{MDC in the
$t$-$t^{\prime}$, $U=8t$ Hubbard model, calculated on a 4-site cluster in
CDMFT. Energy resolution, $\eta=0.1t$ (left and middle). Left: Hole-doped dSC
($t'=-0.3t, n=0.96$), Middle: Electron-doped dSC ($t'=0.3t, n=0.93$), Right:
Same as middle with $\eta=0.02t$. Note the particle-hole transformation in the
electron-doped case. From Ref.~\onlinecite{Kancharla:2005}. }%
\label{fig_75}%
\end{figure}

\subsubsection{CDMFT and DCA}

CDMFT\cite{Kancharla:2005} gives MDC that, at comparable resolution,
$\eta=0.1t$, are again compatible with CPT and with VCPT. The middle panel in
Fig.~\ref{fig_75} is for the electron-doped case but with a particle-hole
transformation so that $t^{\prime}=+0.3t$ and $\mathbf{k\rightarrow k}%
+(\pi,\pi)$. Since there is a non-zero d-wave order parameter in this
calculation, improving the resolution to $\eta=0.02t$ reveals the d-wave gap,
as seen in the right most figure.

\begin{figure}[ptb]
\centerline{\includegraphics[width=8.5cm]{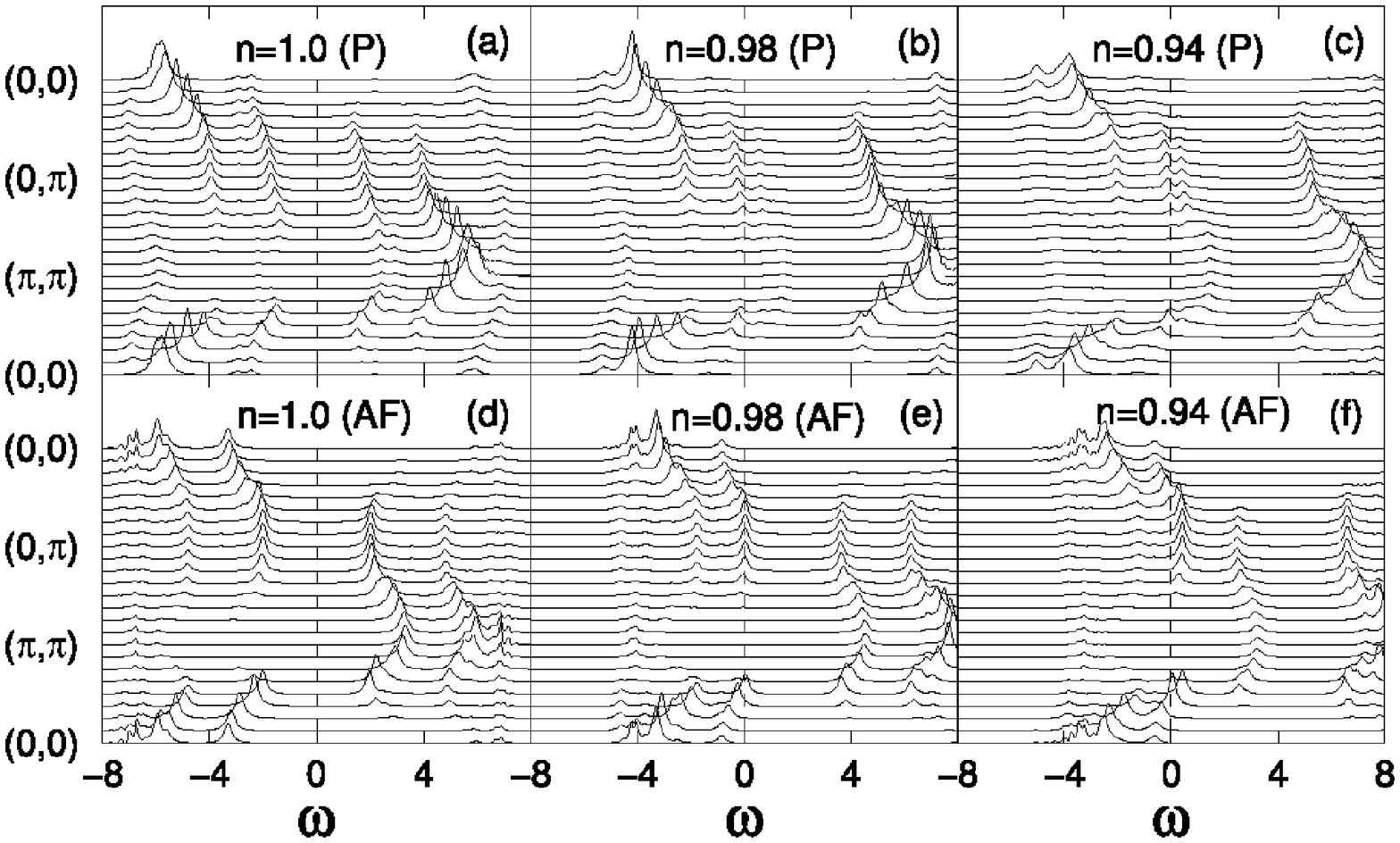}} \caption{EDC in the
$t$-$t^{\prime}=0$, $U=8t$ Hubbard model, calculated on a 4-site cluster in
CDMFT. Top: normal (paramagnetic) state for various densities. Bottom: same
for the antiferromagnetic state. From Ref.~\onlinecite{Kyung:2005}. }%
\label{fig_81}%
\end{figure}

\begin{figure}[ptb]
\centerline{\includegraphics[width=6.5cm]{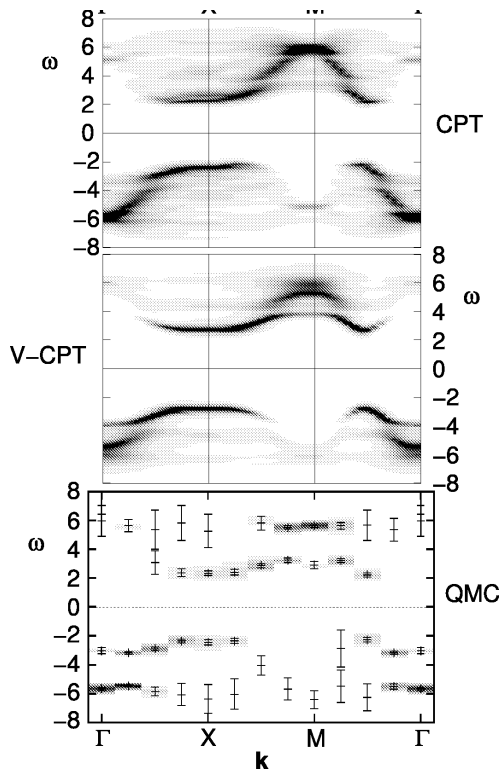}} \caption{EDC in the Hubbard
model, $U=8t$, $t^{\prime}=0$ calculated in CPT, VCPT and QMC. From
Ref.~\onlinecite{Dahnken:2004}. }%
\label{fig_55a}%
\end{figure}

It has been argued for a while in DCA that there is a mechanism whereby
short-range correlations at strong coupling can be the source of the pseudogap
phenomenon.\cite{Huscroft:2001} To illustrate this mechanism in CDMFT, we take
the case $t^{\prime}=t^{\prime\prime}=0$ and compare in Fig.~\ref{fig_81} the
EDC for $U=8t$ without long-range order (top panels) and with long-range
antiferromagnetic order (bottom panels).\cite{Kyung:2005} The four bands
appearing in Figs~\ref{fig_81}a and \ref{fig_81}d are in agreement with what
has been shown\cite{Dahnken:2004, Moreo:1995, Grober:2000} with CPT, VCPT and
QMC in Fig.~\ref{fig_55a}. Evidently there are additional symmetries in the
antiferromagnetic case. The bands that are most affected by the long-range
order are those that are closest to the Fermi energy, hence they reflect spin
correlations, while the bands far from the Fermi energy seem less sensitive to
the presence of long-range order. These far away bands are what is left from
the atomic limit where we have two dispersionless bands. As we dope, the
chemical potential moves into the lower band closest to the Fermi energy. When
there is no long-range order (Figs~\ref{fig_81}b and \ref{fig_81}c) the
lower band closest to the Fermi energy moves very close to it, at the same
time as the upper band closest to the Fermi energy looses weight, part of it
moving closer to the Fermi energy. These two bands leave a pseudogap at the
Fermi energy\cite{Harris:1967, Meinders:1993},
although we cannot exclude that increasing the resolution would
reveal a Fermi liquid at a very small energy scale. In the case when there is
long-range antiferromagnetic order, (Figs~\ref{fig_81}e and \ref{fig_81}f) the
upper band is less affected while the chemical potential moves in the lower
band closest to the Fermi energy but without creating a pseudogap, as if we
were doping an itinerant antiferromagnet. It seems that forcing the spin
correlations to remain short range leads to the pseudogap phenomenon in this
case. When a pseudogap appears, it is created again by very large scattering
rates.\cite{Kyung:2005}

\begin{figure}[ptb]
\centerline{\includegraphics[width=6.5cm]{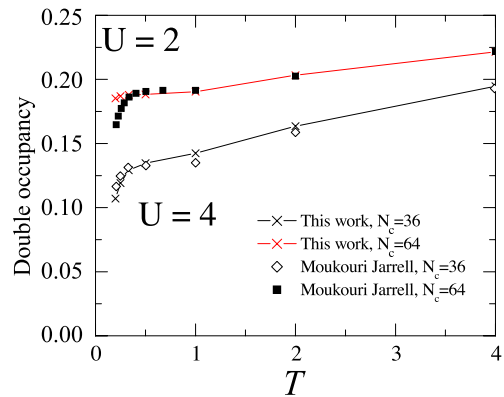}} \caption{ Double occupancy
$\left\langle n_{\uparrow}n_{\downarrow}\right\rangle $, in the
two-dimensional Hubbard model for $n=1$, as calculated from TPSC (lines with
x) and from DCA (symbols) from Ref.~\onlinecite{Moukouri:2001}. Taken from
Ref.~\onlinecite{Kyung:2003a}.}%
\label{fig_83}%
\end{figure}

\subsubsection{TPSC (including analytical results)}

In Hartree-Fock theory, double occupancy is given by $n^{2}/4$ and is
independent of temperature. The correct result does depend on temperature. One
can observe in Fig.~\ref{fig_83} the concordance between the results for the
temperature-dependent double occupancy obtained with DCA and with
TPSC\cite{Kyung:2003a} for the $t^{\prime}=t^{\prime\prime}=0$ model. We have
also done extensive comparisons between straight QMC calculations and
TPSC.\cite{Roy:unpub} The downturn at low temperature has been confirmed by
the QMC calculations. It comes from the opening of the pseudogap due to
antiferromagnetic fluctuations, as we will describe below. The concomitant
increase in the local moment corresponds to the decrease in double-occupancy.
There seems to be a disagreement at low temperature between TPSC and DCA at
$U=2t.$ In fact TPSC is closer to the direct QMC calculation. Since we expect
quantum cluster methods in general and DCA in particular to be less accurate
at weak coupling, this is not too worrisome. At $U=4t$ the density of states
obtained with TPSC and with DCA at various temperatures are very close to each
other.\cite{Kyung:2003a} We stress that as we go to temperatures well below
the pseudogap, TPSC becomes less and less accurate, generally overemphasizing
the downfall in double occupancy.

\begin{figure}[ptb]
\centerline{\includegraphics[width=8.5cm]{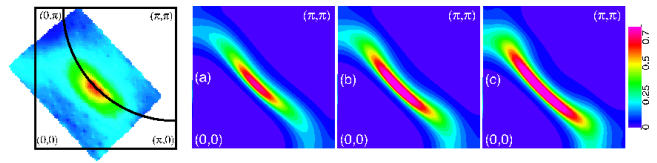}} \caption{MDC at the Fermi
energy for the two-dimensional Hubbard model for $U=6.25$, $t^{\prime
}=-0.175t, t^{\prime\prime}=0.05t$ at various hole dopings, as obtained from
TPSC. The far left from Ref.~\onlinecite{Ronning:2003} is the Fermi surface
plot for $10 \%$ hole-doped Ca$_{2-x}$Na$_x$CuO$_2$Cl$_2$.}%
\label{fig_84}%
\end{figure}
\begin{figure}[ptb]
\centerline{\includegraphics[width=8.5cm]{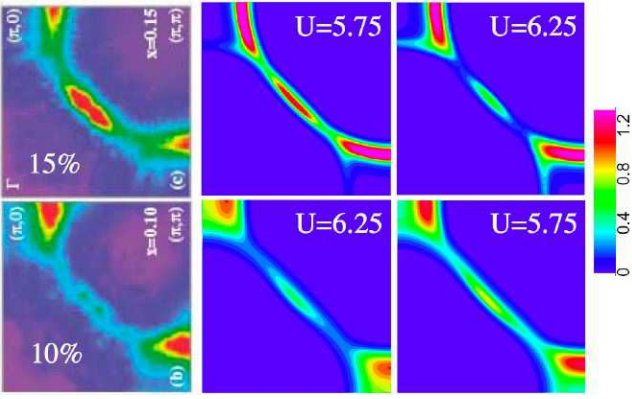}} \caption{MDC at the Fermi
energy in the electron-doped case with $t^{\prime}=-0.175t, t^{\prime\prime
}=0.05t$ and two different $U$'s, $U=6.25t$ and $U=5.75t$ obtained from TPSC.
The first column is the corresponding experimental plots at $10\%$ and $15\%$
doping in Ref.~\onlinecite{Armitage:2002}. From
Refs.~\onlinecite{Kyung:2003,Kyung:2004}.}%
\label{fig_85}%
\end{figure}

We will come back to more details on the predictions of TPSC for the
pseudogap, but to illustrate the concordance with quantum cluster results
shown in the previous subsection, we show in Fig.~\ref{fig_84} MDC obtained at
the Fermi energy in the hole doped case for
$t^{\prime}=-0.175t$, $t^{\prime\prime}=0.05t$. Again there is quasi-particle weight
near $(\pi/2,\pi/2)$ and a pseudogap near $(\pi,0)$ and $(0,\pi)$. However, as
we will discuss below, the antiferromagnetic correlation length necessary to
obtain that pseudogap is too large compared with experiment. The
electron-doped case is shown in Fig.~\ref{fig_85} near optimal doping
and for different values of
$U.$ As $U$ increases, the weight near $(\pi/2,\pi/2)$ disappears. That is in
concordance with the results of CPT shown in Fig.~\ref{fig_64} where the
weight at that location exists only for small $U.$ That also agrees with
slave-boson calculations\cite{Yuan:2005} that found such weight for $U=6t$ and
it agrees also with one-loop calculations\cite{Kusunose:2003} starting from a Hartree-Fock
antiferromagnetic state that did not find weight at that location for $U=8t$.
The simplest Hartree-Fock approach\cite{Kusko:2002, Markiewicz:2004} yields
weight near $(\pi/2,\pi/2)$ only for unreasonably small values of $U$.

\begin{figure}[ptb]
\centerline{\includegraphics[width=6.5cm]{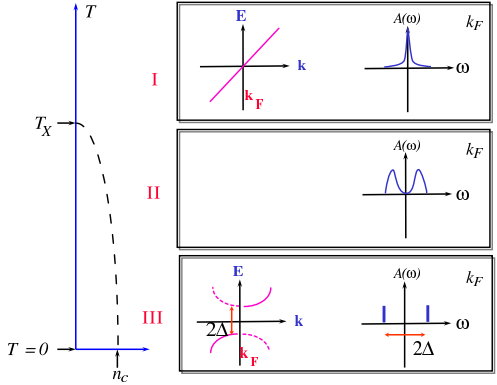}} \caption{Cartoon explanation
of the pseudogap in the weak-coupling limit. Below the dashed crossover line
to the renormalized classical regime, when the antiferromagnetic correlation
length becomes larger than the thermal de Broglie wave length, there appears
precursors of the zero-temperature Boboliubov quasiparticles for the
long-range ordered antiferromagnet.}%
\label{fig_88}%
\end{figure}
A cartoon explanation of the pseudogap is given in Fig.~\ref{fig_88}. At high
temperature we have a Fermi liquid, as illustrated in panel I. Now, suppose we
start from a ground state with long-range order as in panel III, in other
words at a filling between half-filling and $n_{c}$. In the Hartree-Fock
approximation we have a gap and the fermion creation-annihilation operators
now project on Bogoliubov-Valentin quasiparticles that have weight at both
positive and negative energies. In two dimensions, the Mermin-Wagner theorem
means that as soon as we raise the temperature above zero, long-range order
disappears, but the antiferromagnetic correlation length $\xi$ remains large
so we obtain the situation illustrated in panel II, as long as $\xi$ is much
larger than the thermal de Broglie wave length $\xi_{th}\equiv v_{F}/(\pi T)$
in our usual units. At the crossover temperature $T_{X}$ then the relative
size of $\xi$ and $\xi_{th}$ changes and we recover the Fermi liquid. We now
proceed to sketch analytically where these results come from starting from
finite temperature. Details and more complete formulae may be found in
Refs.~\onlinecite{Vilk:1994,Vilk:1996,Vilk:1997,Vilk:1995}. Note also that a
study starting from zero temperature has also been performed in Ref. \onlinecite{borejsza:2004}.

First we show how TPSC recovers the Mermin-Wagner theorem. Consider the
self-consistency conditions given by the local moment sum rule Eq.~(\ref{Suscep})
together with the expression for the spin-susceptibility
Eq.~(\ref{Chi_sp}) and $U_{sp}$ in Eq.~(\ref{ansatz}). First, it is clear that
if the left-hand side of the local moment sum rule Eq.~(\ref{Suscep}) wants to
increase because of proximity to a phase transition, the right-hand side can
do so only by decreasing $\langle n_{\uparrow}n_{\downarrow}\rangle$ which in
turns decreases $U_{sp}$ through Eq.~(\ref{ansatz}) and moves the system away
from the phase transition. This argument needs to be made more precise to
include the effect of dimension. First, using the spectral representation one
can show that every term of $\chi_{sp}({\bf q}, iq_n)$ is positive. Near a phase transition,
the zero Matsubara frequency component of the susceptibility begins to
diverge. On can check from the real-time formalism that the zero-Matsubara
frequency contribution dominates when the characteristic spin fluctuation
frequency $\omega_{sp}\sim\xi^{-2}$ becomes less than temperature, the
so-called renormalized-classical regime. We isolate this contribution on the
left-hand side of the local moment sum rule and we move the contributions from
the non-zero Matsubara frequencies, that are non-divergent, on the right-hand
side. Then, converting the wave vector sum to an integral and expanding the
denominator of the susceptibility around the wave vector where the instability
would occurs to obtain an Ornstein-Zernicke form, the local moment sum rule
Eq.~(\ref{Suscep}) can be written in the form%
\begin{equation}
T\int q^{d-1}dq\frac{1}{q^{2}+\xi^{-2}}=C(T).
\end{equation}
The constant on the right-hand side contains only non-singular contributions
and $\xi^{-2}$ contains $U_{sp}$ that we want to find. From the above
equation, one finds immediately that in $d=2,$ $\xi\approx\exp(C(T)/T)$ so
that the correlation length diverges only at $T=0.$ In three dimensions,
isotropic or not, exponents correspond to those of the $N=\infty$ universality
class.\cite{Dare:1996}

\begin{figure}[ptb]
\centerline{\includegraphics[width=6.5cm]{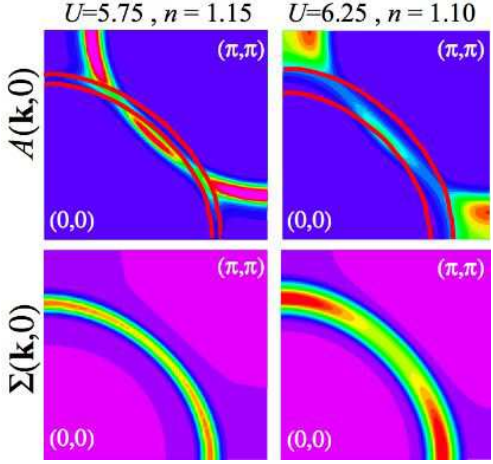}}\caption{MDC plots at the Fermi
energy (upper) and corresponding scattering rates (lower) obtained from TPSC.
The red lines on the upper panel indicate the region where the scattering rate
in the corresponding lower panels is large.}%
\label{fig_91}%
\end{figure}
To see how the pseudogap opens up in the single-particle spectral weight,
consider the expression (\ref{Self-long}) for the self-energy. Normally one
has to do the sum over bosonic Matsubara frequencies first, but the zero
Matsubara frequency contribution has the correct asymptotic behavior in
fermionic frequencies $ik_{n}$ so that one can once more isolate on the
right-hand side the zero Matsubara frequency contribution. This is confirmed
by the real-time formalism\cite{Vilk:1997} (See also Eq.~(\ref{Sigma_Reel})
below). In the renormalized classical regime then, we have
\begin{equation}
\Sigma(\mathbf{k}_{F},ik_{n})\propto T\int q^{d-1}dq\frac{1}{q^{2}+\xi^{-2}%
}\frac{1}{ik_{n}-\varepsilon_{\mathbf{k}_{F}+\mathbf{Q+q}}}
\label{RC-contribution-sigma}%
\end{equation}
where $\mathbf{Q}$ is the wave vector of the instability. Hence, when
$\varepsilon_{\mathbf{k}_{F}+\mathbf{Q}}=0$, in other words at hot spots, we
find after analytical continuation and dimensional analysis that
\begin{align}
\operatorname{Im}\Sigma^{R}(\mathbf{k}_{F},0)  &  \propto-\pi T\int
d^{d-1}q_{\perp}dq_{||}\frac{1}{q_{\perp}^{2}+q_{||}^{2}+\xi^{-2}}\delta
(v_{F}^{\prime}q_{||})\\
&  \propto\frac{\pi T}{v_{F}^{\prime}}\xi^{3-d}.
\label{ImSigma2}
\end{align}
Clearly, in $d=4$, $\operatorname{Im}\Sigma^{R}(\mathbf{k}_{F},0)$ vanishes as
temperature decreases, $d=3$ is the marginal dimension and in $d=2$ we have
that $\operatorname{Im}\Sigma^{R}(\mathbf{k}_{F},0)\propto\xi/\xi_{th}$ that
diverges at zero temperature. In a Fermi liquid that quantity vanishes at zero
temperature. A diverging $\operatorname{Im}\Sigma^{R}(\mathbf{k}_{F},0)$
corresponds to a vanishingly small $A(\mathbf{k}_{F},\omega=0)$ as we can see
from%
\begin{equation}
A(\mathbf{k},\omega)=\frac{-2\operatorname{Im}\Sigma^{R}(\mathbf{k}_{F}%
,\omega)}{(\omega-\varepsilon_{\mathbf{k}}-\operatorname{Re}\Sigma
^{R}(\mathbf{k}_{F},\omega))^{2}+\operatorname{Im}\Sigma^{R}(\mathbf{k}%
_{F},\omega)^{2}}. \label{Spectral-weight_1}%
\end{equation}
To see graphically this relationship between the location of the pseudogap and
large scattering rates at the Fermi surface, we draw in Fig.~\ref{fig_91} both
the Fermi surface MDC and, in the lower panels, the corresponding plots for
$\operatorname{Im}\Sigma^{R}(\mathbf{k},0)$. Note that at stronger $U$ the
scattering rate is large over a broader region, leading to a depletion of
$A(\mathbf{k,}\omega)$ over a broader range of $\mathbf{k}$ values.

An argument for the splitting in two peaks seen in Figs.~\ref{fig_29} and
\ref{fig_88} is as follows. Consider the singular renormalized contribution
coming from the spin fluctuations in Eq.~(\ref{RC-contribution-sigma}) at
frequencies $\omega\gg v_{F}\xi^{-1}.$ Taking into account that contributions
to the integral come mostly from a region $q\leq\xi^{-1}$, this expression
leads to%
\begin{align}
\operatorname{Re}\Sigma^{R}( \mathbf{k}_{F},\omega)  &  =\left(  T\int
q^{d-1}dq\frac{1}{q^{2}+\xi^{-2}}\right)  \frac{1}{ik_{n}-\varepsilon
_{\mathbf{k}_{F}+\mathbf{Q}}}\nonumber\\
&  \equiv\frac{\Delta^{2}}{\omega-\varepsilon_{\mathbf{k}_{F}+\mathbf{Q}}}%
\end{align}
which, when substituted in the expression for the spectral weight
(\ref{Spectral-weight_1}) leads to large contributions when
\begin{equation}
\omega-\varepsilon_{\mathbf{k}}-\frac{\Delta^{2}}{\omega-\varepsilon
_{\mathbf{k}_{F}+\mathbf{Q}}}=0
\end{equation}
or, equivalently,%
\begin{equation}
\omega=\frac{( \varepsilon_{\mathbf{k}}+\varepsilon_{\mathbf{k}_{F}%
+\mathbf{Q}}) \pm\sqrt{( \varepsilon_{\mathbf{k}}-\varepsilon_{\mathbf{k}%
_{F}+\mathbf{Q}}) ^{2}+4\Delta^{2}}}{2},
\end{equation}
which corresponds to the position of the hot spots in Fig.~\ref{fig_85} for example.

Note that analogous arguments hold for any fluctuation that becomes
soft,\cite{Vilk:1997} including superconducting ones.\cite{Allen:2001,
Kyung:2001}
The wave vector $\mathbf{Q}$ would be different in each case.

\subsubsection{Weak- and strong-coupling pseudogaps}

The CPT results of Figs.~\ref{fig_62} and \ref{fig_71} clearly show that the
pseudogap is different from the Mott gap. At finite doping, the Mott gap
remains a local phenomenon, in the sense that there is a region in frequency
space that is not tied to $\omega=0$ where \textit{for all wave vectors }there
are no states. The peudogap by contrast is tied to $\omega=0$ and occurs in
regions nearly connected by $(\pi,\pi),$ whether we are talking about hole- or
about electron-doped cuprates. That the phenomenon is caused by short-range
correlations can be seen in CPT from the fact that the pseudogap is
independent of cluster shape and size (most of the results were presented for
$3\times4$ clusters and we did not go below size $2\times2$). The
antiferromagnetic correlations and any other two-particle correlations do not
extend beyond the size of the lattice in CPT. Hence, the pseudogap phenomenon
cannot be caused by antiferromagnetic long-range order since no such order
exists in CPT. This is also vividly illustrated by the CDMFT results in
Fig.~\ref{fig_81} that contrast the case with and without antiferromagnetic
long-range order. The CDMFT results also suggest that the pseudogap appears in
the bands that are most affected by antiferromagnetic correlations hence it
seems natural to associate it with short-range spin correlations. The value of
$t^{\prime}$ has an effect, but it mostly through the fact that it has a
strong influence on the relative location of the antiferromagnetic zone
boundary and the Fermi surface, a crucial factor determining where the
pseudogap is. All of this as well as many results obtained earlier by
DCA\cite{Huscroft:2001} suggest that there is a strong coupling mechanism that
leads to a pseudogap in the presence of only short-range two-body
correlations. However, the range cannot be zero. Only the Mott gap appears at
zero range, thus the pseudogap is absent in single-site DMFT.

In the presence of a pseudogap at strong coupling ($U>8t$), wave vector is not, so to speak, such a bad quantum number in certain directions. In other words the wave description is better in those directions.
In other directions that are ``pseudogapped'', it is as if the localized, or particle description was better.
This competition between wave and particle behavior is inherent to the Hubbard model.
At the Fermi surface in low dimension, it seems that this competition is resolved by dividing (it is a crossover, not a real division) the Fermi surface in different sections.

There is also a weak-coupling mechanism for the pseudogap. This has been
discussed at length just in the previous section on TPSC. Another way to
rephrase the calculations of the previous section is in the real frequency
formalism. There one finds\cite{Vilk:1997} that
\begin{align}
&\Sigma^{\prime\prime R}(\mathbf{k}_{F},\omega)\nonumber\\
&~~ \propto\int\frac{d^{d-1}%
q_{\perp}}{(2\pi)^{d-1}}\int\frac{d\omega^{\prime}}{\pi}[n(\omega^{\prime
})+f(\omega+\omega^{\prime})]\chi_{sp}^{\prime\prime}(\mathbf{q}%
;\omega^{\prime}) \label{Sigma_Reel}%
\end{align}
so that if the characteristic spin fluctuation frequency in $\chi_{sp}%
^{\prime\prime}(\mathbf{q};\omega^{\prime})$ is much larger than temperature,
then $[n(\omega^{\prime})+f(\omega+\omega^{\prime})]$ can be considered to act
like a window of size $\omega$ or $T$ and $\chi_{sp}^{\prime\prime}%
(\mathbf{q};\omega^{\prime})$ can be replaced by a function of ${\bf q}$ times
$\omega^{\prime}$ which immediately leads to the Fermi liquid result
$[\omega^{2}+(\pi T)^{2}].$ In the opposite limit where the characteristic
spin fluctuation frequency in $\chi_{sp}^{\prime\prime}(\mathbf{q}%
;\omega^{\prime})$ is much less than temperature, then it acts as a window
narrower than temperature and $[n(\omega^{\prime})+f(\omega+\omega^{\prime})]$
can be approximated by the low frequency limit of the Bose factor, namely
$T/\omega^{\prime}.$ Using the thermodynamic sum rule, that immediately leads
to the result discussed before in Eq.(\ref{ImSigma2}),
$\operatorname{Im}\Sigma(\mathbf{k}%
_{F},0)\propto(\pi T/v_{F}^{\prime})\xi^{3-d}.$ This mechanism for the
pseudogap needs long correlation lengths. In CPT, this manifests itself by the
fact that the apparent pseudogap in Fig.~\ref{fig_64} at $U=4t$ is in fact
mostly a depression in spectral weight that depends on system size and shape.
In addition, in contrast to the short-range strong-coupling mechanism, at weak
coupling the pseudogap is more closely associated with the intersection of the
antiferromagnetic zone boundary with the Fermi surface.

Which mechanism is important for the cuprates will be discussed below in the
section on comparisons with experiments.

\subsection{d-wave superconductivity}

The existence of d-wave superconductivity at weak coupling in the Hubbard
model mediated by the exchange of antiferromagnetic fluctuations
\cite{Scalapino:1995, Carbotte:1999} had been proposed even before the
discovery of high-temperature superconductivity.\cite{Scalapino:1986,
Miyake:1986, Beal-Monod:1986} At strong-coupling, early
papers\cite{Kotliar:1988, Inui:1988} also proposed that the superconductivity
would be d-wave. The issue became extremely controversial, and even recently
papers have been published\cite{Zhang:1997} that suggest that there is no
d-wave superconductivity in the Hubbard model. That problem could have been
solved very long ago through QMC calculations if it had been possible to do
them at low enough temperature. Unfortunately, the sign problem hindered these
simulations, and the high temperature
results\cite{Hirsch:1988,White:1989,Moreo:1991,Scalettar:1991} were not
encouraging: the d-wave susceptibility was smaller than for the
non-interacting case. Since that time, numerical results from variational
QMC,\cite{Paramekanti:2001, Paramekanti:2004} exact
diagonalization\cite{Poilblanc:2002} and other numerical
approaches\cite{Sorella:2002} for example, suggest that there is indeed d-wave
superconductivity in the Hubbard model.

In the first subsection, we show that VCPT leads to a zero-temperature phase
diagram for both hole and electron-doped systems that does show the basic
features of the cuprate phase diagram, namely an antiferromagnetic phase and a
d-wave superconducting phase in doping ranges that are quite close to
experiment\cite{Senechal:2005} (The following section will treat in more
detail comparisons with experiment). The results are consistent with
CDMFT.\cite{Kancharla:2005} The fall in the d-wave superconducting order
parameter near half-filling is associated with the Mott phenomenon. The next
subsection stresses the instability towards d-wave superconductivity as seen
from the normal state and mostly at weak coupling. We show that TPSC can
reproduce available QMC results and that its extrapolation to lower
temperature shows d-wave superconductivity in the Hubbard model. The
transition temperature found at optimal doping\cite{Kyung:2003} for $U=4t$
agrees with that found by DCA,\cite{Maier:2005a} a result that could be
fortuitous. But again the concordance between theoretical results obtained at
intermediate coupling with methods that are best at opposite ends of the range
of coupling strengths is encouraging.

\begin{figure}[ptb]
\centerline{\includegraphics[width=8cm]{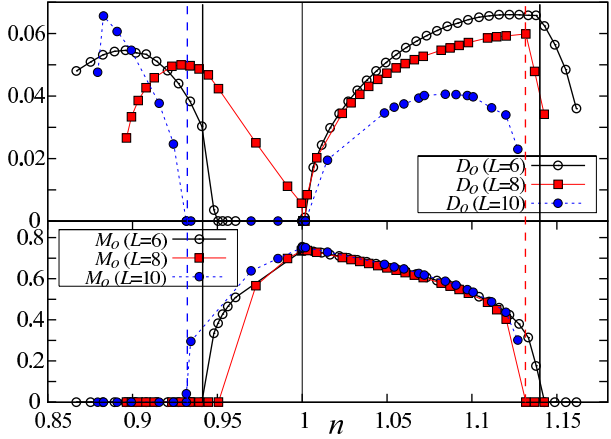}} \caption{Antiferromagnetic
(bottom) and $d$-wave (top) order parameters for $U=8t$, $t^{\prime}=-0.3t$
$t^{\prime\prime}=0.2t$ as a function of the electron density ($n$) for
$2\times3$, $2\times4$ and 10-site clusters, calculated in VCPT. Vertical
lines indicate the first doping where only $d$-wave order is non-vanishing.
From Ref.~\onlinecite{Senechal:2005}. }%
\label{fig_96}%
\end{figure}

\begin{figure}[ptb]
\centerline{\includegraphics[width=7cm]{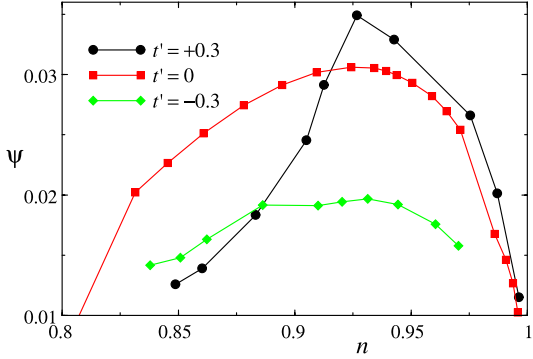}} \caption{d-wave order parameter
as a function of $n$ for various values of $t^{\prime}$, calculated in CDMFT
on a $2 \times2$ cluster for $U=8t$. The positive $t^{\prime}$ case
corresponds to the electron-doped case when a particle-hole transformation is
performed. From Ref.~\onlinecite{Kancharla:2005}. }%
\label{fig_138}%
\end{figure}

\subsubsection{Zero-temperature phase diagram}

In VCPT, one adds to the cluster Hamiltonian the terms\cite{Senechal:2005}
\begin{align}
H'_{M} &  =M\sum_{\mu}e^{i\mathbf{Q}\cdot\mathbf{R}_{\mu}}(n_{\mu\uparrow
}-n_{\mu\downarrow})\label{weiss_eq}\\
H'_{D} &  =D\sum_{\mu\nu}g_{\mu\nu}(c_{\mu\uparrow}c_{\nu\downarrow
}+\mathrm{H.c.})
\end{align}
with $M$ and $D$ are respectively antiferromagnetic and d-wave superconducting
Weiss fields that are determined self-consistently and
$g_{\mu\nu}$ equal to $\pm1$ on near-neighbor sites following the d-wave pattern.
We recall that the cluster Hamiltonian should be understood in a variational
sense. Fig.~\ref{fig_96} summarizes, for various cluster sizes, the results
for the d-wave order parameter $D_{0}$ and for the antiferromagnetic order
parameter $M_{0}$ for a fixed value of $U=8t$ and the usual hopping parameters
$t^{\prime}=-0.3t$ and $t^{\prime\prime}=0.2t$. The results for
antiferromagnetism are quite robust and extend over ranges of dopings that
correspond quite closely to those observed experimentally. Despite the fact
that the results for d-wave superconductivity still show some size dependence,
it is clear that superconductivity alone without coexistence extends over a
much broader range of dopings on the hole-doped than on the electron-doped
side as observed experimentally. VCPT calculations on smaller system
sizes\cite{Hanke:2005} but that include, for thermodynamic consistency, 
the cluster chemical potential as a
variational parameter show superconductivity that extends over a much broader
range of dopings. Also, for small $2\times2$ clusters, VCPT has stronger order
parameter on the electron than on the hole-doped side, contrary to the results
for the largest system sizes in Fig.~\ref{fig_96}. This is also what is found
in CDMFT as can be seen in Fig.~\ref{fig_138}. It is quite likely that the
zero-temperature Cooper pair size is larger than two sites, so we consider the
results for $2\times2$ systems only for their qualitative value.

Concerning the question of coexistence with antiferromagnetism, one can see
that it is quite robust on the electron-doped side whereas on the hole-doped
side, it is size dependent. That suggests that one should also look at
inhomogeneous solutions on the hole-doped side since stripes are generally
found experimentally near the regions where antiferromagnetism and
superconductivity meet.

\begin{figure}[ptb]
\centerline{\includegraphics[width=8cm]{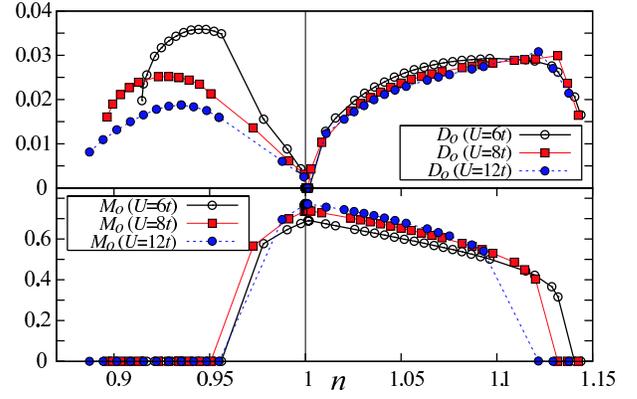}} \caption{Antiferromagnetic
(bottom) and $d$-wave (top) order parameters as a function of the electron
density ($n$) for $t^{\prime}=-0.3t$ $t^{\prime\prime}=0.2t$ and various
values of $U$ on a 8-site cluster, calculated in VCPT. From
Ref.~\onlinecite{Senechal:2005}. }%
\label{fig_97}%
\end{figure}

\begin{figure}[ptb]
\centerline{\includegraphics[width=7cm]{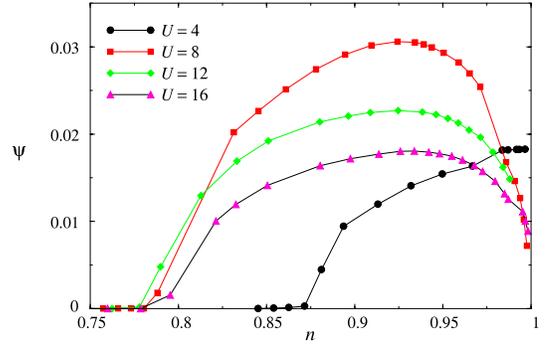}} \caption{d-wave order parameter
as a function of $n$ for various values of $U$, and $t^{\prime}=t^{\prime
\prime}=0$ calculated in CDMFT on a 4-site cluster. From
Ref.~\onlinecite{Kancharla:2005}. }%
\label{fig_139}%
\end{figure}

Fig.~\ref{fig_97} shows clearly that at strong coupling the size of the order
parameter seems to scale with $J$, in other words it decreases with increasing
$U.$ This is also found in CDMFT,\cite{Kancharla:2005} as shown in Fig.
\ref{fig_139} for $t^{\prime}=t^{\prime\prime}=0$.

\begin{figure}[ptb]
\centerline{\includegraphics[width=6.5cm]{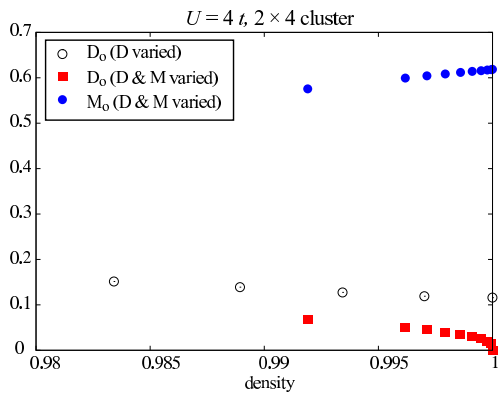}} \caption{VCPT calculations for
$U=4t, t^{\prime}=t^{\prime\prime}=0$ near half-filling on $2 \times4$
lattice. Contrary to the strong coupling case, the d-wave order parameter $D_0$
survives all the way to half-filling at weak coupling, unless we also allow
for antiferromagnetism.}%
\label{fig_1}%
\end{figure}

If we keep the antiferromagnetic order parameter to zero, one can check with
both VCPT and CDMFT (Fig.~\ref{fig_138}) that the d-wave superconducting order
parameter goes to zero at half-filling. This is clearly due to Mott
localization. Indeed, at smaller $U=4t$ for example, the order parameter does
not vanish at half-filling if we do not allow for long-range antiferromagnetic
order, as illustrated in Fig.~\ref{fig_139} for CDMFT.\cite{Kancharla:2005}
The same result was found in VCPT, as shown in Fig.\ref{fig_1}.\cite{Senechal:2005} 
Restoring long-range
antiferromagnetic order does however make the d-wave order parameter vanish at half-filling.

There are thus two ways to make d-wave superconductivity disappear at
half-filling, either through long antiferromagnetic correlation lengths\cite{Micnas:2005} 
or
through the Mott phenomenon. In the real systems, that are Mott insulators and
also antiferromagnets at half-filling, both effects can contribute.

\subsubsection{Instability of the normal phase}

In the introduction to this section, we alluded to QMC calculations for the
d-wave susceptibility.\cite{Hirsch:1988,White:1989,Moreo:1991,Scalettar:1991}
Recent results\cite{Kyung:2003} for that quantity as a function of doping for
various temperatures and for $U=4t,$ $t^{\prime}=t^{\prime\prime}=0$ are shown
by symbols in Fig.~\ref{fig_12}. For lower temperatures, the sign problem
prevents the calculation near half-filling. Yet, the lowest temperature is low
enough that a dome shape begins to appear. Nevertheless, comparison with the
non-interacting case, shown by the top continuous line, leads one to believe
that interactions only suppress d-wave superconductivity. We can easily
understand why this is so.
As we already know,
the TPSC results obtained from Eq.~(\ref{Suscep_d}) are very close to the QMC
calculations, as shown by the solid lines in Fig.~\ref{fig_12}.
In the temperature range of interest, the main
contribution comes from the first term in Eq.~(\ref{Suscep_d}). That term
represents the contribution to the susceptibility that comes from dressed
quasiparticles that do not interact with each other. Since dressed
quasiparticles have a lifetime, a pair breaking effect, it is normal that this
contribution to the interacting susceptibility leads to a smaller contribution
than in the non-interacting case. At the lowest temperature, $\beta=4/t$, the
vertex contribution represented by the second term in Eq.~(\ref{Suscep_d})
accounts for about $20\%$
of the total. It goes in the direction of increasing the susceptibility. As we
decrease the temperature further in TPSC, that term becomes comparable with
the first one. Since the vertex in Eq.~(\ref{Suscep_d}) accounts for the
exchange of a single spin wave, equality with the first term signals the
divergence of the series, as in $1/(1-x)\sim1+x$. The divergence of that
series represents physically the instability of the normal phase to a d-wave
superconducting phase. This is analogous to the Thouless criterion and hence
it gives an upper bound to $T_{c}$. In other words,
Berezinskii-Kosterlitz-Thouless physics is not included.

\begin{figure}[ptb]
\centerline{\includegraphics[width=6.5cm]{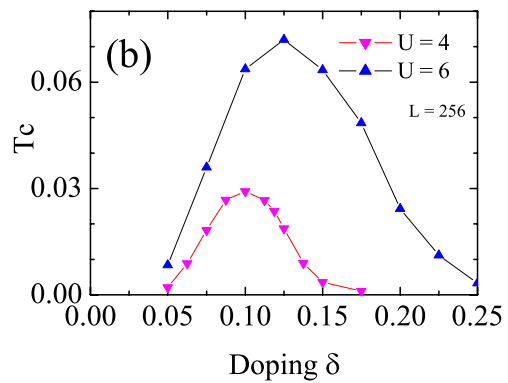}} \caption{$T_{c}$ as a
function of doping, $\delta=1-n$, for $t^{\prime}=t^{\prime\prime}=0$
calculated in TPSC using the Thouless criterion. From
Ref.~\onlinecite{Kyung:2003}.}%
\label{fig_104}%
\end{figure}

Fig.~\ref{fig_104} shows the TPSC transition temperature for $U=4t$ and for $U=6t$.
As we move towards half-filling, located to the left of the diagram, starting
from large dopings, $T_{c}$ goes up because of the increase in
antiferromagnetic fluctuations. Eventually, $T_{c}$ turns around and decreases
because of the opening of the pseudogap. Physically, when the pseudogap opens
up, weight is removed from the Fermi level, the density of states becomes very
small, and pairing cannot occur any more. In the FLEX
approximation\cite{Bickers:1989, Pao:1994} that does not exhibit a
pseudogap,\cite{Deisz:1996} that downturn, observed already in QMC at high
temperature, does not occur. We have observed that in cases where $t^{\prime
}\neq0$ so that the pseudogap opens only in a limited region around hot spots,
the downturn can become less pronounced.

The case $n=0.9=1-\delta$ that corresponds to optimal doping for $U=4t$ in
Fig.~\ref{fig_104} has been studied by DCA. In an extensive and systematic
study of the size dependence, Maier \textit{et al. }\cite{Maier:2005a}
established the existence of a d-wave instability at a temperature that
coincides to within a few percent with the result in Fig.~\ref{fig_104}. Since
very few vortices can fit within even the largest cluster sizes studied in
Ref.~\onlinecite{Maier:2005a}, it is quite likely that the $T_{c}$ that they find
does not include Berezinskii-Kosterlitz-Thouless effects, just like ours.
Despite the fact that, again, the concordance between weak and strong coupling
methods at intermediate coupling comforts us, the uncertainties in the results
found with TPSC and DCA force us to also allow for a fortuitous coincidence.

\section{Comparisons with experiment\label{Experiment}}

The reduction of the real problem of high-temperature superconducting
materials to a one-band Hubbard is a non-trivial one. It has been discussed
already in the early days of high $T_{c}$ superconductivity. The notion of a
Zhang-Rice singlet\cite{Zhang:1988a} emerged for hole doped systems. The
mapping to a one-band model has been discussed in many
references,\cite{Andersen:1995, Macridin:2005} and we do not wish to discuss
this point further here. In fact it is far from obvious that this mapping is
possible. It is known that about $0.5$ eV below the Fermi surface, that
mapping fails in hole-doped systems.\cite{Macridin:2005} Nevertheless, the
one-band Hubbard model is in itself a hard enough problem for us. So it is
satisfying to see that, in the end, it gives a picture that agrees with
experiment in a quite detailed manner for the ARPES spectrum near the Fermi
surface, for the phase diagram as well as for neutron scattering in cases
where it can be calculated.

Although we will not come back on this point at all, we briefly mention that
fitting the spin wave spectrum\cite{Coldea:2001} for all energies and wave vectors at half
filling in La$_{2}$CuO$_{4}$ gives values of $U,t,t^{\prime},t^{\prime\prime}$
that are close to those used in the rest of this paper.\cite{Delannoy:2005,
Delannoy:2005a, Toader:2005, Raymond:2005}
It is in this context that ring exchange terms are usually discussed.

\subsection{ARPES spectrum, an overview\label{ARPES_overview}}

ARPES experiments have played a central role in the field of high-temperature
superconductivity. We cannot expect to be able to present the vast
experimental literature on the subject. We refer the reader to a very
exhaustive review\cite{Damascelli:2003} and to some less complete but recent
ones.\cite{Damascelli:2003a, Kordyuk:2005} The main facts about ARPES have
been summarized in Ref.~\onlinecite{Damascelli:2003}. We comment on their main
points one by one, using italics for our paraphrase of the reported
experimental observations.

\textit{(i) The importance of Mott Physics and the renormalization of the
bandwidth from $t$ to $J$ for the undoped parent compounds}. This
renormalization was clear already in early QMC
calculations.\cite{Preuss:1995,Moreo:1995} We already discussed the presence
of four peaks. The one nearest to the Fermi surface at negative energies is
the one referred to by experimentalists when they refer to this
renormalization. This band has a dispersion of order $J$ (not shown on
Fig.~\ref{fig_55a},
but see Ref. \onlinecite{Preuss:1995}). This result also agrees with
VCPT as shown in Fig.~\ref{fig_55a} and CDMFT (Fig.~\ref{fig_81}a). As shown
in Figs.~\ref{fig_81}a and \ref{fig_81}d, whether the state is ordered or not
the band width is similar. Analytical strong-coupling
expansions\cite{Pairault:1998, Pairault:2000} and exact diagonalizations also
find the same result. To find detailed agreement with experiment, one needs to
include $t^{\prime}$ and $t^{\prime\prime}$.\cite{Gooding:1994} The evolution
of the position of chemical potential for extremely small dopings as discussed
in Ref.~\onlinecite{Shen:2005} is not reproduced by the strong-coupling calculations,
although the result on chemical potential is somewhat material 
dependent.\cite{Yoshida:2005}

\textit{(ii) In the overdoped case, the Fermi surface is well defined.}
Although we have not shown any figures concerning this point, VCPT and CDMFT
show the same result.

\begin{figure}[ptb]
\centerline{\includegraphics[width=5cm]{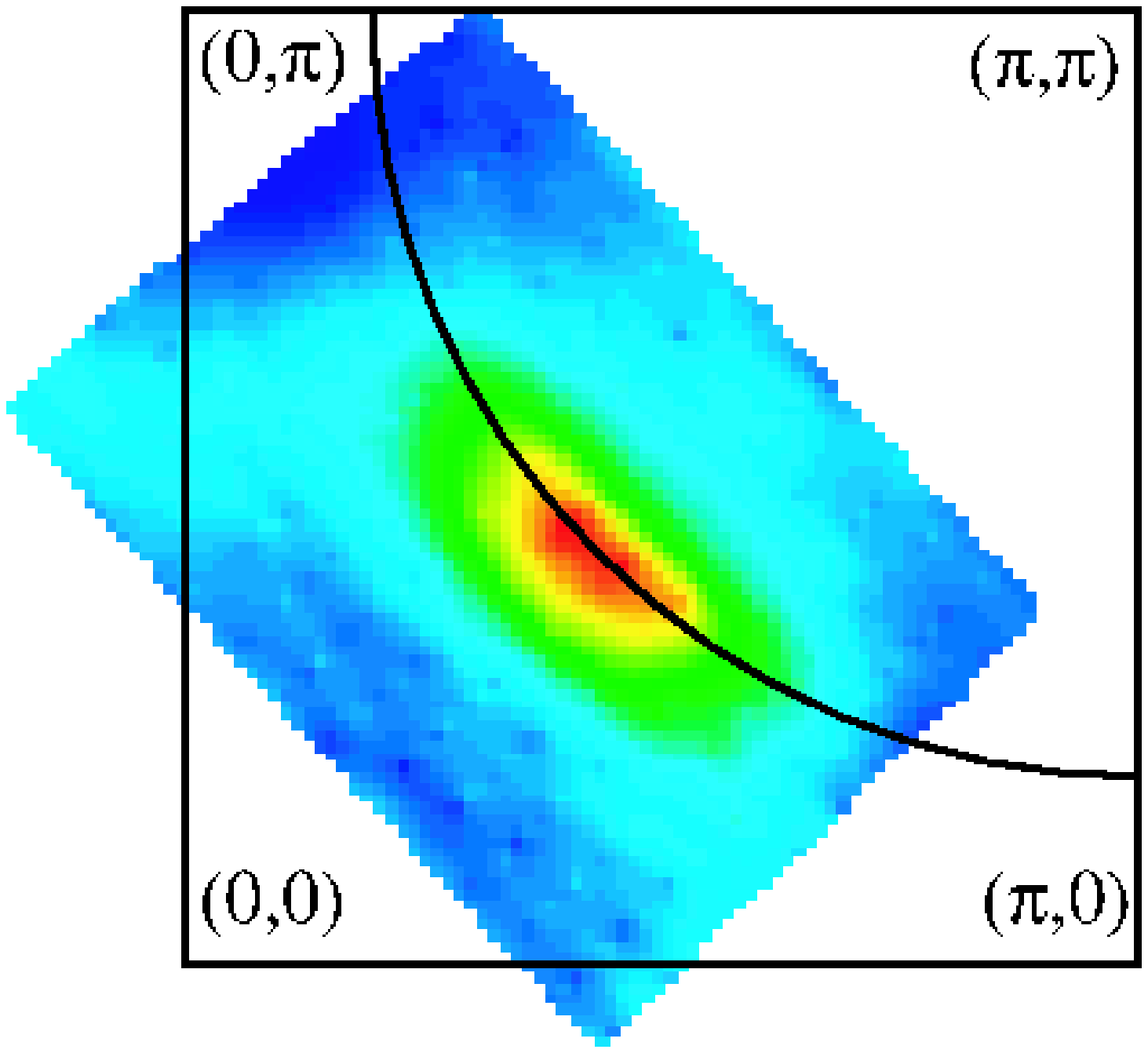}}\caption{MDC at the Fermi energy
for $10 \%$ hole-doped Ca$_{2-x}$Na$_x$CuO$_2$Cl$_2$ from
Ref.~\onlinecite{Ronning:2003}.}%
\label{fig_107}%
\end{figure}

\textit{(iii) The evolution with doping of the electronic structure has been
mapped. It has shown the importance of antiferromagnetic correlations in the
p-type underdoped cuprates and especially in the n-type ones in which the
hot-spot physics is still observed at optimal doping. } We will come back on
the latter point for electron-doped cuprates in the following subsection. The
strong-coupling results obtained with VCPT and CDMFT have a resolution of
order $0.1t$, which translates into about $30$~meV. This is not enough to
accurately measure the Fermi velocity, which was found to be doping
independent in LSCO.\cite{Zhou:2003} However, this suffices to compare with
MDC curves obtained experimentally by integrating over an energy range of
about $60$~meV, as shown in Fig.~\ref{fig_107} obtained in
Ref.~\onlinecite{Ronning:2003} on Calcium oxyclorate Ca$_{2-x}$Na$_{x}%
$CuO$_{2}$Cl$_{2}$, a $10\%$ hole-doped high temperature superconductor. The
similarities between that figure and the CPT (Fig.~\ref{fig_64}), VCPT
(Fig.~\ref{fig_110}) and CDMFT (Fig.~\ref{fig_75}) results is striking. The
agreement is better when no antiferromagnetic long-range order is assumed, as
in the CPT case. The flattening of the band structure near $(\pi,0)$ observed
experimentally, can also be seen in CPT by comparing the top and middle EDC's
taken at small and large $U$ respectively on the left panel of
Fig.~\ref{fig_63}. This flattening is associated with the pseudogap
phenomenon. Recall that the theoretical results were obtained with $t^{\prime
}=-0.3t$ and $t^{\prime\prime}=0.2t$. This in turn implies an electron-hole
asymmetry that is observed experimentally. We come back to this in the
following subsection.

\textit{(iv) The overall d-wave symmetry of the superconducting gap has been
observed for both hole and electron doping, supporting the universality of the
pairing nature in the cuprates. } In the next to next subsection, we discuss
the phase diagram for competing antiferromagnetism and d-wave
superconductivity and show striking similarities with the observations.

\textit{(v) A normal-state pseudogap has been observed to open up at a
temperature }$T^{\ast}>T_{c}$ \textit{in the underdoped regime with a d-wave
form similar to the one of the superconducting gap. }That statement is correct
only in the hole-doped compounds. In electron-doped systems the pseudogap has
a form that is not of d-wave shape. If $T_{c}$ comes from a universal pairing
mechanism, a universal mechanism may also be behind the pseudogap. As we have
already discussed however, there are quantitative differences between strong
and weak coupling mechanisms for both $T_{c}$ and the pseudogap. For
electron-doped systems, we made quantitative predictions for the value of
$T^{\ast}$ that have later been confirmed experimentally. All this is
discussed further below. To date, in cluster methods the pseudogap temperature
has been studied only with DCA.\cite{Huscroft:2001}

\textit{(vi) A coherent quasiparticle peak below }$T_{c}$\textit{ has been
observed near }$(\pi,0)$\textit{ whose spectral weight scales with the doping
level }$x$\textit{ in the underdoped regime. } We expect that it is a general
result that long-range order will restore quasiparticle like excitations in
strongly correlated systems because gaps remove scattering channels near the
Fermi level. This is clearly illustrated by comparing the upper and lower
panels in Fig.~\ref{fig_81} that contrast the same spectra with and without
antiferromagnetic long-range order. We have not performed the analysis of our
results yet that could tell us whether the spectral weight of the
quasiparticle scales with $x$ in the hole-underdoped regime. Our resolution
may not be good enough to see the quasiparticle peak. Sharpening of the
quasiparticle excitations in the superconducting state has however been
observed in DCA.\cite{Maier:2004}

\textit{(vii) The presence of an energy scale of about }$40-80$~meV \textit{in
the quasiparticle dynamics manifests itself through a sharp dispersion
renormalization and drop in the scattering rate observed at those energies at
different momenta. } In hole-doped systems there is a kink in the nodal
direction that is already seen above $T_{c}$ while in the antinodal direction
it appears only below $T_{c}$. The energy scales and doping dependences of
these two kinks are also different.\cite{Gromko:2003} The energy resolution in
VCPT and CDMFT is not sufficient to distinguish these subtleties. In
electron-doped cuprates experiments\cite{Armitage:2003} suggest that there is
no observable kink feature, in agreement with the results presented in the
following subsection.

\subsection{The pseudogap in electron-doped cuprates}

\begin{figure}[ptb]
\centerline{\includegraphics[width=6.5cm]{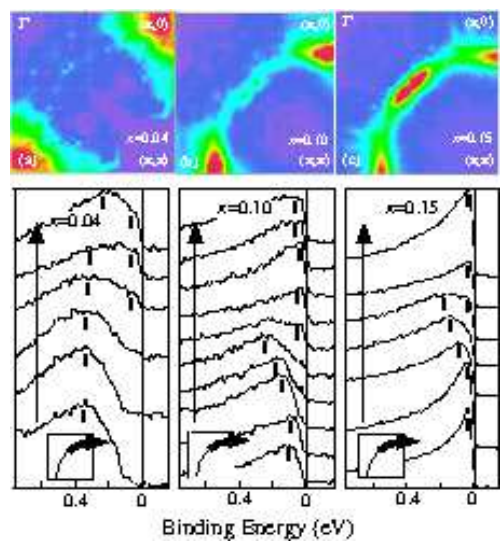}} \caption{Doping dependence
of the MDC from experiments on NCCO with the corresponding EDC. From
Ref.~\onlinecite{Armitage:2002} }%
\label{fig_121a}%
\end{figure}

The ARPES spectrum of electron-doped cuprates is strikingly different from
that of their hole-doped counterpart. The Fermi energy MDC's for the first
quadrant of the Brillouin zone\cite{Armitage:2002} are shown at the top of
Fig.~\ref{fig_121a} for three different dopings. There is a very clear
evolution with doping. At the lowest dopings, there is no weight near
$(\pi/2,\pi/2),$ contrary to the hole-doped case shown in Fig.~\ref{fig_107}.
For all dopings there is weight near $(\pi,0)$ instead of the pseudogap that
appeared there in the hole-doped case. The EDC's, also shown on the bottom of
Fig.~\ref{fig_121a}, are drawn for a trajectory in the Brillouin zone that
follows what would be the Fermi surface in the non-interacting case. Regions
that are more green than red on the corresponding MDC's along that trajectory
are referred to as hot spots. On the EDC's we clearly see that hot spots do
not correspond to simply a decrease in the quasiparticle weight $Z$. They
truly originate from a pseudogap, in other words from the fact that the
maximum is pushed away from zero energy. Even though the measurements are done
at low temperature $(T=10-20K)$ the energy resolution of about $60$~meV makes
the superconducting gap invisible. What is observed at this resolution is the pseudogap.

\begin{figure}[ptb]
\centerline{\includegraphics[width=8.5cm]{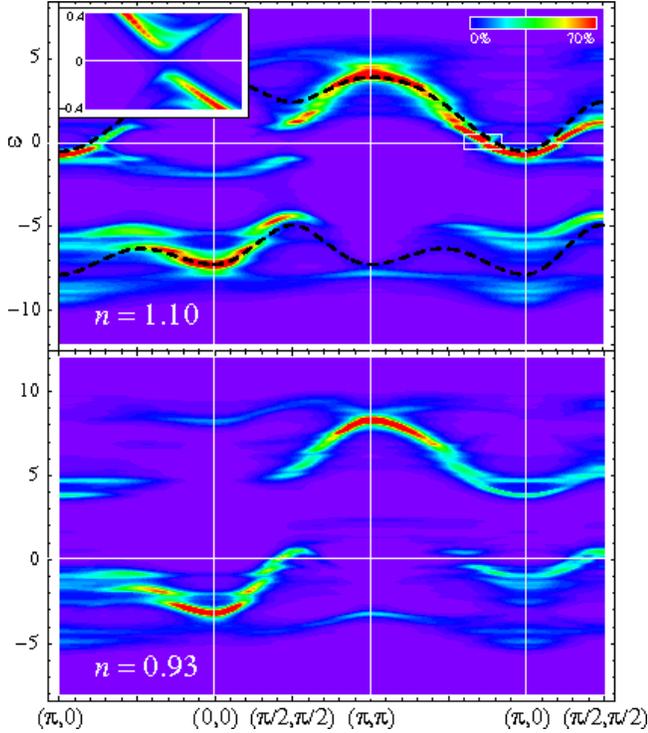}} \caption{Intensity plot of
the spectral function as a function of $\omega$ in units of $t$ and wave
vector from VCPT for $U=8t$ $t^{\prime}=-0.3t, t^{\prime\prime}=0.2t$ and
$n=0.93$ at the bottom and $n=1.10$ (electron-doped) at the top. The
Lorentzian broadening is $0.12t$ in the main figure and $0.04t$ in the inset
that displays the d-wave gap. Top panel is for the electron-doped case in the
right-hand panel of Fig.~\ref{fig_110}, while bottom panel is for the
hole-doped case on the left of Fig.~\ref{fig_110}. From
Ref.~\onlinecite{Senechal:2005}.}%
\label{fig_117}%
\end{figure}

\begin{figure}[ptb]
\centerline{\includegraphics[width=6.5cm]{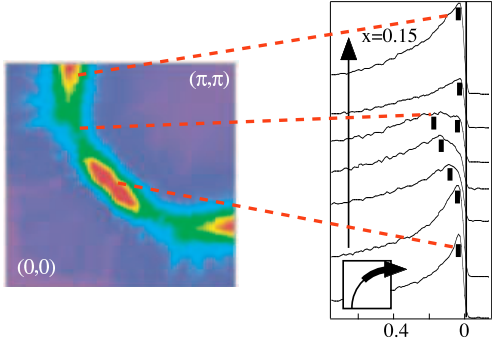}} \caption{Experimental Fermi
surface plot (MDC at the Fermi energy) for NCCO (left) and corresponding
energy distribution curves (right) for $15 \%$ electron-doping. From
Ref.~\onlinecite{Armitage:2002} }%
\label{fig_121}%
\end{figure}

The contrast between the location of the pseudogap in the hole and
electron-doped compounds is clearly seen in Fig.~\ref{fig_117} obtained from
VCPT.\cite{Senechal:2005} In that figure, the magnitude of the spectral weight
is represented by the different colors as a function of frequency (in units of
$t$) along different cuts of the Brillouin zone. In the bottom panel, for the
hole-doped case, one observes the pseudogap near $(\pi,0)$. In the top panel,
for the electron-doped case, it is only by zooming (inset) on the region for the Fermi
energy crossing near $(\pi,0)$ that one sees the d-wave superconducting gap.
At $10\%$ electron-doping, the pseudogap near $(\pi/2,\pi/2)$ is apparent. In
this case there is antiferromagnetic long-range order, but even if we use CPT
that does not exhibit long-range order, there appears a pseudogap in that
region.\footnote{The latter result is disputed by a recent DCA calculation by
Alexandru Macridin, Mark Jarrell, Thomas Maier, P. R. C. Kent,
cond-mat/0509166.} The main difference between CPT and VCPT results is the
bending back of the bands (for example around the symmetry axis at $(\pi
/2,\pi/2)$) caused by halving of the size of the Brillouin zone in the
antiferromagnetic case. Form factors\cite{Kusko:2002} are such that the
intensity is not symmetric even if the dispersion is. The faint band located
at an energy about $t$ below the Fermi energy near $(\pi/2,\pi/2)$ was also
found in Ref.~\onlinecite{Kusunose:2003} by a one-loop spin-wave calculation
around the Hartree-Fock antiferromagnetic ground state at $U=8t$.
Experimentalists\cite{Armitage:2002} have suggested the existence of these
states. The VCPT results go well beyond the spin-wave calculation (dashed
lines in Fig.~\ref{fig_117}) since one can also see numerous features in
addition to remnants of the localized atomic levels around $+5t$ and $-10t$.

The optimally doped case is the real challenge for strong-coupling
calculations. The spin-wave approach in Ref.~\onlinecite{Kusunose:2003} never
shows the weight near $(\pi/2,\pi/2)$ that is seen in experiment
(Fig.~\ref{fig_121}). Early mean-field calculations by Kusko et
al.\cite{Kusko:2002} suggest that this $(\pi/2,\pi/2)$ feature appears for
$U=3t$. This is very small compared with $U$ of the order of the bandwidth
$8t$ necessary to have a Mott insulator at half-filling. We already discussed
in Sec. \ref{WeakAndStrongCoupling} that both CPT and TPSC show that weight
near $(\pi/2,\pi/2)$ appears for $U$ not too large, say of order $6t$. This
same result is also obtained in the Kotliar-Ruckenstein slave boson
approach.\cite{Yuan:2005}

\begin{figure}[ptb]
\centerline{\includegraphics[width=6.5cm]{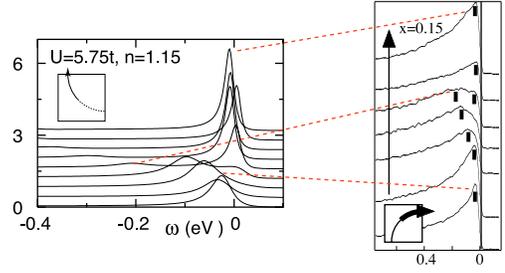}}\caption{EDC $A_{<}(\mathbf{k},\omega)\equiv A(\mathbf{k},\omega)f(\omega)$ along the Fermi surface
calculated in TPSC (left) at optimal doping for $t^{\prime}=-0.175t$,
$t^{\prime\prime}=0.05t$, $t=350$~meV and corresponding ARPES data on NCCO
(right). From Ref.~\onlinecite{Kyung:2004}.}%
\label{fig_122}%
\end{figure}

\begin{figure}[ptb]
\centerline{\includegraphics[width=6.5cm]{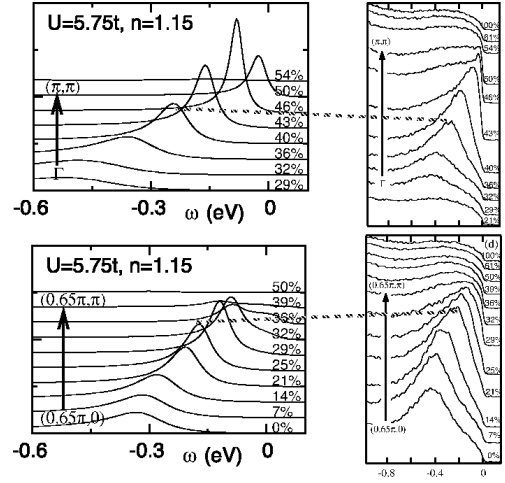}}\caption{EDC $A_{<}(\mathbf{k},\omega)\equiv A(\mathbf{k},\omega)f(\omega)$ along two other directions
calculated for $t^{\prime}=-0.175t$, $t^{\prime\prime}=0.05t$, $t=350$~meV in
TPSC (left column) and corresponding ARPES data on NCCO (right column). From
Ref.~\onlinecite{Kyung:2004}.}%
\label{fig_123}%
\end{figure}

Since TPSC is valid for a system of infinite size, we present detailed
comparisons\cite{Kyung:2004} with experiment\cite{Armitage:2002} on
Nd$_{1.85}$Ce$_{0.15}$CuO$_{4}$, an electron-doped cuprate. We take
$t^{\prime}=-0.175t$, $t^{\prime\prime}=0.05t.$ Results obtained with
$t^{\prime}=-0.275$ are very close to those we present. With the values used
in CPT, $t^{\prime}=-0.3t,$ $t^{\prime}=0.2t,$ $U=6t,$ TPSC does not lead to
strong enough antiferromagnetic fluctuations to obtain non-trivial effects in
the temperature range studied, $\beta=20t$. We take $t=350$~meV.
Fig.~\ref{fig_121a} shows the correspondence between EDC and MDC. Comparisons
with experimental EDC at wave vectors along the non-interacting Fermi surface
appear in Fig.~\ref{fig_122} for $U=5.75t$ and $15\%$ doping $(n=1.15)$. The
dashed lines indicate the quite detailed agreement between theory and
experiment. At the hot spot, (middle dashed line), the weight is pushed back
about $0.2$~eV and there is a very small peak left at the Fermi surface, as in
the experiment. If $U$ is not large enough the antiferromagnetic fluctuations
are not strong enough to lead to a pseudogap. As in CPT (Fig.~\ref{fig_64}), if
$U$ is too large the $(\pi/2,\pi/2)$ weight disappears, as illustrated earlier
in Fig.~\ref{fig_85}. In Fig.~\ref{fig_123}, cuts along the $(0,0)$ to
$(\pi,\pi)$ and $(0.65\pi,0)$ to $(0.65\pi,\pi)$ directions are compared with
experiment. Again the peak positions and widths are very close, except for
some experimental tails extending in the large binding energy direction. The
theoretical results have similar asymmetry, but not as pronounced.
Experimentally, the large binding energy tails (\textquotedblleft the
background\textquotedblright) are the least reproducible features from sample to sample, especially for wave vectors near the Fermi
 surface".
\footnote{Kyle Shen, Private communication.} The experimental
renormalized Fermi velocities are $3.31\times10^{5}$~m/s and $3.09\times
10^{5}$~m/s along the zone diagonal and along the $(\pi,0)$-$(\pi,\pi)$
direction, respectively. The corresponding renormalized Fermi velocities
obtained by TPSC are $3.27\times10^{5}$~m/s and $2.49\times10^{5}$~m/s,
respectively. The agreement is very good, particularly along the diagonal
direction. The bare Fermi velocities are renormalized in TPSC by roughly a
factor of two.\cite{Hankevych:2005}

\begin{figure}[ptb]
\centerline{\includegraphics[width=6.5cm]{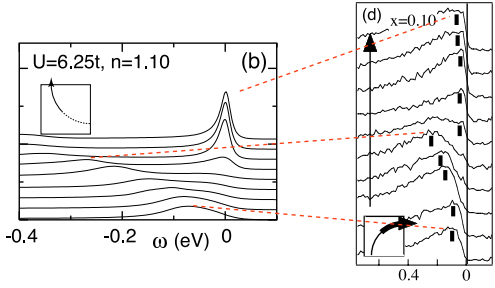}} \caption{EDC $A_{<}(\mathbf{k},\omega)\equiv A(\mathbf{k},\omega)f(\omega)$ along the Fermi surface shown in
the insets for (b) $n=1.10$, $U=6.25t$. Lines are shifted by a constant for
clarity. From Ref.~\onlinecite{Kyung:2004}. }%
\label{fig_124}%
\end{figure}

As we move towards half-filling, we have to increase $U$ slightly to find
agreement with experiment, as discussed earlier in Fig.~\ref{fig_85}.
Fig.~\ref{fig_124} shows how well the EDC's agree for a Fermi surface cut at
$10\%$ doping $(n=1.10)$. The increase is expected physically from the fact
that with fewer electrons the contribution to screening that comes from Thomas
Fermi physics should not be as good. This is also consistent with the fact
that a larger value of $U$ is necessary to explain the Mott insulator at
half-filling. It would also be possible to mimic the ARPES spectrum by keeping
$U$ fixed and changing the hopping parameters, but the changes would be of
order $20\%,$ which does not appear realistic.\cite{Kyung:2004}

\begin{figure}[ptb]
\centerline{\includegraphics[width=5.5cm]{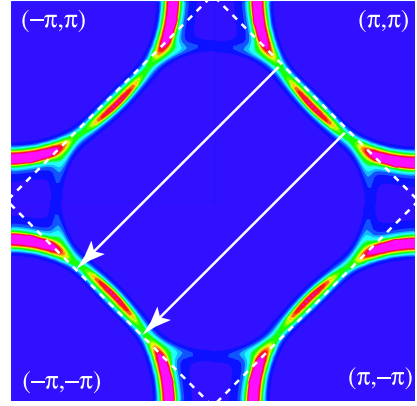}}\caption{Hot spots from
quasi-static scatterings off antiferromagnetic fluctuations (renormalized
classical regime).}%
\label{fig_125}%
\end{figure}
\begin{figure}[ptb]
\centerline{\includegraphics[width=6.5cm]{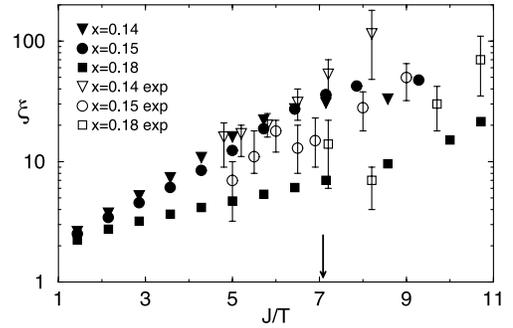}} \caption{Semi-log plot of the
AFM correlation length (in units of the lattice constant) against inverse
temperature (in units of $J=125$ meV). Filled symbols denote calculated
results and empty ones experimental data of Ref.~\onlinecite{Mang:2004} and
Ref.~\onlinecite{Matsuda:1992} ($x=0.15$). From Ref.~\onlinecite{Kyung:2004}.
}%
\label{fig_126}%
\end{figure}

We have already explained that the physics behind the pseudogap in TPSC is
scattering by nearly critical antiferromagnetic fluctuations. This is
illustrated in Fig.~\ref{fig_125}. If this explanation is correct, the
antiferromagnetic correlation length measured by neutron scattering should be
quite large. The results of the measurement\cite{Mang:2004, Matsuda:1992} and
of the TPSC calculations are shown in Fig.~\ref{fig_126}. The agreement is
again surprisingly good. As we move to smaller dopings $n=1.1$ (not shown) the
agreement becomes less good, but we do expect TPSC to deteriorate as $U$
increases with underdoping. The arrow points to the temperature where EDC's
shown earlier were calculated. Note however that the neutron measurements were
done on samples that were not reduced, by contrast with the ARPES measurements
mentioned earlier. We are expecting experiments on this subject.\footnote{M.
Greven, private communication.}. We should point out that the EDC's depend
strongly on temperature and on the actual value of the antiferromagnetic
correlation length only in the vicinity of the temperature where there is a
crossover to the pseudogap regime.  Decreasing the temperature makes the
$\omega=0$ peaks near $(\pi,0)$ sharper\cite{Kyung:2004}
as observed experimentally.\cite{Onose:2004}

\begin{figure}[ptb]
\centerline{\includegraphics[width=6.5cm]{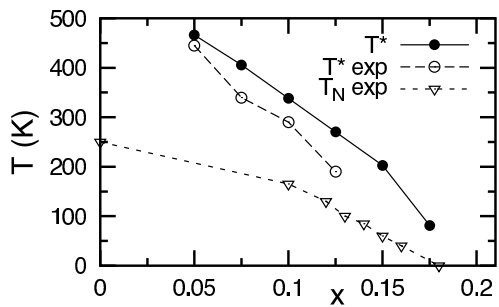}} \caption{Pseudogap
temperature $T^{\ast}$ (filled circles denote $T^{\ast}$ calculated from TPSC,
empty ones experimental data extracted from optical
conductivity.\cite{Onose:2001}) Empty triangles are experimental N\'{e}el
temperatures $T_{N}$. The samples are reduced.\cite{Mang:2004} From
Ref.~\onlinecite{Kyung:2004}. }%
\label{fig_127}%
\end{figure}

The ARPES pseudogap temperature $T^{\ast}$ has been predicted with
TPSC.\cite{Kyung:2004} The predictions are shown by the solid line in
Fig.~\ref{fig_127}. The pseudogap temperature observed in optical
experiments\cite{Onose:2001} is shown by the open circles. It differs from the
ARPES result, especially as we move towards optimal doping. The size of the
pseudogap observed in the optical experiments\cite{Onose:2001} $(10T^{\ast})$
is comparable to the ARPES pseudogap. The solid line in Fig.~\ref{fig_127}
contains several predictions. If we look at $13\%$ doping $(n=1.13)$, the line
predicts $T^{\ast}\sim250K$. Experiments that were done without being aware of
this prediction\cite{Matsui:2005} have verified it. It would be most
interesting to do neutron scattering experiments on the same samples to check
whether the antiferromagnetic correlation length $\xi$ and the thermal de
Broglie wave length $\xi_{th}$ are comparable at that temperature, as
predicted by TPSC. Fig.~\ref{fig_127} also predicts that the pseudogap induced
by antiferromagnetic fluctuations will disappear at the quantum critical point
where long-range antiferromagnetic order disappears, in other words it will
coincide with the crossing of the experimentally observed N\'{e}el temperature
(dashed line with triangles in Fig.~\ref{fig_127}) with the zero temperature
axis (if that crossing is not masked by the superconducting transition).
Recent optical conductivity experiments\cite{Millis:2004, Lobo:2005} confirm
this prediction as well.

In TPSC, superconducting fluctuations can also lead to a pseudogap by an
analogous mechanism.\cite{Vilk:1997}

\begin{figure}[ptb]
\centerline{\includegraphics[width=6.5cm]{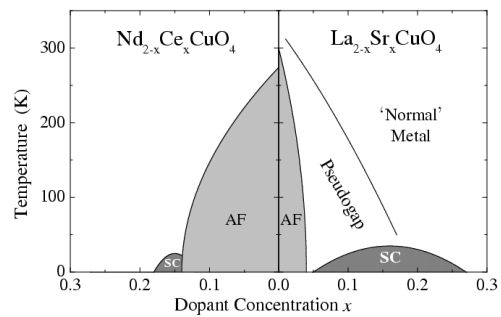}} \caption{The generic phase
diagram of high-$T_{c}$ superconductors, from
Ref.~\onlinecite{Damascelli:2003}. There should also be a pseudogap line on
the electron-doped side. It was not well studied at the time of publication of
that paper.}%
\label{fig_133}%
\end{figure}

\subsection{The phase diagram for high-temperature superconductors}

The main features appearing in the phase diagram of high-temperature
superconductors are the pseudogap phase, the antiferromagnetic phase and the
d-wave superconducting phase. Fig.~\ref{fig_133}\cite{Damascelli:2003} shows
the typical diagram with hole doping to the right and electron doping to the
left. Zero on the horizontal axis corresponds to half-filling. There are other
features on the phase diagram, in particular checkerboard
patterns\cite{Hanaguri:2004} or stripe phases\cite{Stock:2004} that appear in
general close to the region where antiferromagnetism and superconductivity
come close to each other. Before we try to understand these more detailed
features, one should understand the most important phases. In the previous
subsection we have discussed the pseudogap phase, in particular on the
electron-doped side (not indicated on Fig. \ref{fig_133}). A recent review of
the pseudogap appears in Ref. \onlinecite{Norman:2005}. In the following, we
discuss in turn the phase diagram and then the nature of the superconducting phase
itself and its relation to the Mott phenomenon.

\begin{figure}[ptb]
\centerline{\includegraphics[width=6.5cm]{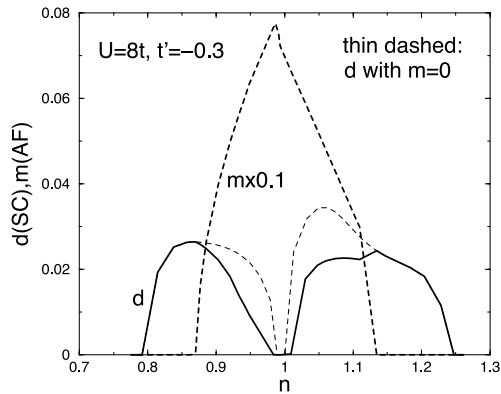}}\caption{Antiferromagnetic
order parameter $m$ (dashed) and d-wave (solid) order parameter obtained from
CDMFT on a $2 \times2$ cluster. The result obtained by forcing $m=0$ is also
shown as a thin dashed line.}%
\label{fig_136}%
\end{figure}

\subsubsection{Competition between antiferromagnetism and superconductivity}

We have already shown in Fig.~\ref{fig_96} the prediction of VCPT for the
zero-temperature phase diagram.\cite{Senechal:2005} Here, we just point out
how closely the position of the antiferromagnetic phase boundary, appearing in
the lower panel, coincides with the experimental phase diagram in
Fig.~\ref{fig_133}. (Note that electron concentration increases from right to
left on this experimental phase diagram). In particular, there is little size
dependence to the position of this boundary, (6 to 10 sites) and in addition
the dependence on the value of $U$ is also weak, as can be seen from
Fig.~\ref{fig_97}. Hence, the positions of the antiferromagnetic phase
boundaries is a robust prediction of VCPT. The CDMFT result for a four site
cluster in a bath is shown in Fig.~\ref{fig_136} for $t'=-0.3t,t^{\prime
\prime}=0$ and $U=8t$. The agreement with experiment is not as good. Despite
the useful presence of a bath in CDMFT, the cluster itself is of size
$2\times2,$ which is probably smaller than the Cooper pair size. We can obtain results closer to those of VCPT by increasing the variational space.

The d-wave superconducting order parameter on the top panel of
Fig.~\ref{fig_96} shows more size dependence than the antiferromagnetic order
parameter. Nevertheless, there are some clear tendencies: (a) d-wave
superconductivity can exist by itself, without antiferromagnetism. The
vertical lines indicate the location of the end of the antiferromagnetic phase
for the various system sizes to help this observation. (b) The range where
d-wave-superconductivity exists without antiferromagnetism, is about three
times larger on the hole than on the electron-doped side, as observed
experimentally. (c) As system size increases, the maximum d-wave order
parameter is larger on the hole than on the electron-doped side. (d) The
tendency to have coexisting antiferromagnetism and d-wave superconductivity is
rather strong on the electron-doped side of the phase diagram. This is
observed experimentally\cite{Fournier:1998} but only over a rather narrow
region near optimal doping. Recent experiments\cite{Motoyama:unpub} challenge
this result, others\cite{Sonier:2003, Sonier:2004} indicate that
antiferromagnetism can be induced from the d-wave superconducting phases with
very small fields. (e) On the hole-doped side, d-wave superconductivity and
antiferromagnetism coexist for a very narrow range of dopings for system size
$N_{c}=6$, for a broad range extending to half-filling for $N_{c}=8$ and not
at all for $N_{c}=10$. In other words, the tendency to coexistence is not even
monotonic. We interpret this result as a reflection of the tendency to form
stripes observed experimentally on the hole-doped side.\cite{Stock:2004,
Wakimoto:2000, Wakimoto:2001} We cannot study systems large enough to allow
for striped inhomogeneous states to check this statement.

\begin{figure}[ptb]
\centerline{\includegraphics[width=6.5cm]{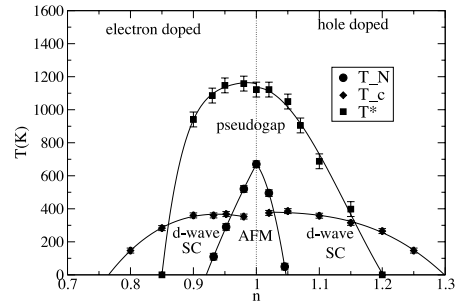}} \caption{Phase diagram
obtained from DCA for $U=8t$ for the two-band model. From
Ref.~\onlinecite{Macridin:2005}}%
\label{fig_136b}%
\end{figure}

The more realistic two-band model has also been studied using
DCA.\cite{Macridin:2005} The results are shown on Fig. \ref{fig_136b}.
Electron concentration increases from right to left. This phase diagram is
very close to that obtained with the same method from the one-band Hubbard
model\cite{Macridin:2005} with $t^{\prime}=-0.3t,$ $t''=0$,
$U=8t$. The qualitative results agree with the other calculations and with
experiment: antiferromagnetism extends over a narrower doping range for hole
than for electron doping and d-wave superconductivity by itself exists over a
broader range for the hole-doped case than for the electron-doped case. The
actual ranges where antiferromagnetism and d-wave superconductivity exist are
not in as good an agreement with experiment as in the VCPT case. However, as
in CDMFT, the system sizes, $2\times2$, are very small. Overall then, quantum
cluster methods, VCPT in particular, allow us to obtain from the Hubbard model
the two main phases, antiferromagnetic and d-wave superconducting, essentially
in the observed doping range of the zero-temperature phase diagram. At finite
temperature, DCA and TPSC agree on the value of $T_{c}$ for the particle-hole
symmetric model at $10\%$ doping and $U=4t$. Recent studies of the irreducible
vertex using DCA\cite{Maier:2005b} also show that in the weak-coupling
limit the particle-particle d-wave
channel leads to an instability driven by antiferromagnetic fluctuations as
temperature decreases, as found in TPSC.

To understand the effect of pressure on the phase diagram, note that $U/t$
should decrease as pressure increases since the increase in the overlap
between orbitals should lead mainly to an increase in $t$. Hence, as can be
deduced from Fig.~\ref{fig_104}, applying pressure should lead to a decrease
in the value of $T_{c}$ at weak coupling, concomitant with the decrease in
antiferromagnetic fluctuations that lead to pairing in the weak coupling case.
This is indeed what pressure does experimentally in the case of electron-doped
high-temperature superconductors,\cite{Maple:1990} reinforcing our argument
that near optimal doping they are more weakly coupled. It is
widely known on the other hand that pressure \textit{increases} $T_{c}$ in
hole-doped systems. That is consistent with the strong-coupling result that we
found in VCPT and CDMFT, namely that the maximum d-wave order parameter in
that case scales with $J=4t^{2}/U,$ a quantity that increases with $t$ and
hence pressure. Whereas in the weak coupling case superconductivity is a
secondary phenomenon that occurs after antiferromagnetic fluctuations have
built up, in strong coupling they can be two distinct phenomena as can be seen
from the phase diagram, even though they arise from the same microscopic
exchange interaction represented by $J$.

\subsubsection{Anomalous superconductivity near the\ Mott transition}

Superconductivity in the underdoped regime is very much non-BCS. First of all,
we notice in Fig.~\ref{fig_139} obtained in CDMFT\cite{Kancharla:2005} that at
strong coupling the d-wave superconducting order parameter vanishes as we move
towards half-filling even in the absence of long-range antiferromagnetic
order. In other words, the Mott phenomenon by itself suffices to destroy
d-wave superconductivity. This conclusion is reinforced by the fact that at
weak coupling ($U=4t$) where there is no Mott localization, d-wave
superconductivity survives at half-filling. In the presence of
antiferromagnetic long-range order, that last statement would not be true, as
confirmed by VCPT calculations in Fig.~\ref{fig_1}: at $U=4t$ d-wave
superconductivity survives at half-filling if we do not allow for
antiferromagnetic long-range order but it disappears if we do. In BCS theory,
the presence of an interaction $J$ that leads to attraction in the d-wave
channel would lead at $T=0$ to d-wave superconductivity at all dopings
including half-filling, unless we allow for competing long-range order. At
strong coupling, no long-range order is necessary to destroy d-wave superconductivity.

Superconductivity at strong coupling\cite{Haslinger:2003, Marsiglio:2005} also
differs from BCS in the origin of the condensation energy. Suppose we do BCS
theory on the attractive Hubbard model. Then, as in the usual BCS model,
kinetic energy is increased in the superconducting state because the Fermi
surface is no-longer sharp. On the other hand, in the superconducting phase
there is a gain in potential energy. The reverse is true at strong coupling. This result follows from DCA\cite{Maier:2004}
and is in agreement with the kinetic energy drop in the superconducting state
that has been estimated from the \textit{f-}sum rule in optical conductivity
experiments.\cite{Santander-Syro:2003, Molegraaf:2002,Deutscher:2005}
Photoemission data\cite{Norman:2000} had also suggested this kinetic energy
drop in the superconducting state. A crossover from non-BCS-like to BCS
behavior in the condensation mechanism as we go from underdoping to overdoping
has also been seen recently experimentally.\cite{Deutscher:2005} We do not
seem to have the resolution to find that crossover since the condensation
energy becomes very small on the overdoped side. We expect that crossover from
strong to weak coupling will also lead to a change from a kinetic-energy
driven to a potential-energy driven pairing mechanism. This is confirmed by
CDMFT calculations for the attractive Hubbard model.\cite{Kyung:2005a}

\begin{figure}[ptb]
\centerline{\includegraphics[width=6.5cm]{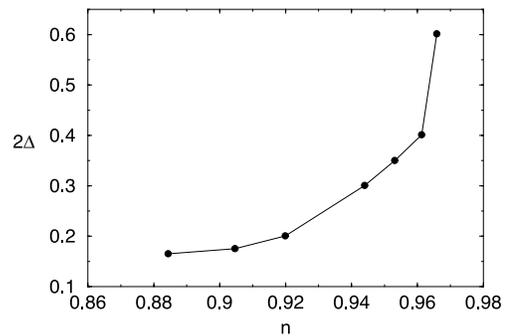}}\caption{The gap
in the density of states of the dSC as a function of filling for $U=8t$, $t^{\prime}=-0.3t$ as calculated in CDMFT on a
$2 \times2$ cluster. From Ref.~\onlinecite{Kancharla:2005}.}%
\label{fig_141}%
\end{figure}
A third way in which superconductivity in the underdoped regime is non-BCS is
that the drop in the order parameter as we go towards half-filling is
accompanied by an increase in the gap as measured in the
single-particle density of states. Fig.~6 of Ref.~\onlinecite{Sutherland:2003}
summarizes the experimental evidence for the increase in the size of the
gap. That increase, observed in the CDMFT calculation of the
gap, is illustrated in Fig.~\ref{fig_141}.\cite{Kancharla:2005} That gap has essentially the same size as that observed in the normal pseudogap state\cite{Kyung:2005}.

\section{Conclusion, open problems\label{Conclusion}}

High-temperature superconductivity has forced both experimentalists and
theorists to refine their tools and to develop new ones to solve the puzzles
offered by this remarkable phenomenon. From a theoretical perspective, the
original suggestion of Anderson\cite{Anderson:1987} that the physics was in
the one-band Hubbard model is being confirmed. In the absence of \textit{ab
initio} methods to tell us what is the correct starting point, such insight is
essential. The non-perturbative nature of the phenomenon has however forced
theorists to be extremely critical of each other's theories since none of them
can pretend that a small parameter controls the accuracy of the approximations.

If theorists are to convince each other and experimentalists that a solution
of the high-temperature superconductivity problem has been found, then the
theories have to give quantitative results and to make predictions. Unlike
most traditional problems in condensed matter physics however, the
non-perturbative nature of the problem means that no simple mean-field like
theory can be trusted, even if it seems to agree qualitatively with
experiment. In fact several such theories have been
proposed\cite{Kotliar:1988, Inui:1988, Anderson:2005, Bickers:1987} not long after the
experimental discovery of the phenomenon but they have not been accepted
immediately. Theories have to be internally consistent, they have to agree
with exact results whenever they are available, and then they can be compared
with experiments. If there is a disagreement with experiment, the starting
point (one-band Hubbard model) needs to be reconsidered. When approaches
developed on the basis of weak-coupling ideas agree at intermediate coupling
with approaches developed on the basis of strong-coupling ideas, then one
gains confidence in the validity of the results. We have argued that such
concordance is now found in a number of cases and that corresponding rather
detailed quantitative agreement with experiment can be found. In a
non-perturbative context it becomes essential to also cross check various approaches.

The main theoretical methods that we have discussed are those that we have
developed or perfected or simply used in our group: The Two-Particle
Self-Consistent approach that is based on weak-coupling but non-perturbative
ideas (no diagrams are involved), as well as heavily numerical approaches such
as QMC and various quantum cluster methods, VCPT and CDMFT.

Based on our own work and that of many others, we think the following
experimental facts about high-temperature superconductivity can be reproduced
very accurately by calculations for the one-band Hubbard model with $U$ in the
intermediate coupling range ($U\sim8t$) with $t\sim350$~meV, and hopping
parameters $t^{\prime}$ and $t^{\prime\prime}$ close to the values suggested
by band structure calculations,\cite{Andersen:1995} namely $t^{\prime}=-0.3t,$
$t^{\prime\prime}=0.2t$.

(i) In the one-band Hubbard model the main phases of the zero-temperature
phase diagram, namely antiferromagnetic and d-wave superconducting, appear
very near the observed ranges for both the hole- and electron-doped cases.

(ii) The normal state is unstable to a d-wave superconducting phase in a
temperature range that has the correct order of magnitude. As usual the value
of $T_{c}$ is the most difficult quantity to evaluate since one must take into
account Kosterlitz-Thouless physics as well as the effect of higher dimensions
etc, so this level of agreement must be considered satisfying.

(iii) The ARPES MDC at the Fermi energy and the EDC near the Fermi energy are
qualitatively well explained by cluster calculations for both hole- and
electron-doped cases. These comparisons, made at a resolution of about $30$ to
$60$ meV are not very sensitive to long-range order, although order does
influence the results. One is mainly sensitive to the pseudogap, so this is
the main phenomenon that comes out from the model. Energy resolution is not
good enough to see a kink. More details about what aspects of ARPES are
understood may be found in Sec. \ref{ARPES_overview}.

(iv) In the case of electron-doped cuprates, the value of $U$ near optimal
doping seems to be in the range $U\sim6t$, which means that it is accessible
to studies with TPSC that have better resolution. In that case, the agreement
with experiment is very accurate, even if there is room for improvement and a need for further experiments. In
addition, the value of $T^{\ast}$ for 13\% doping has been predicted
theoretically before it was observed experimentally, one of the very rare
predictions in the field of high-temperature superconductors. All of this
agreement with ARPES data is strong indication that $U\sim6t$ is appropriate
to describe electron-doped superconductors near optimal doping. Additional
arguments come from the pressure dependence of the superconducting transition
temperature $T_{c},$ which increases with $t/U$ contrary to the
strong-coupling result, and from simple ideas on Thomas-Fermi screening. The
latter would predict that the screened interaction scales like $\partial
\mu/\partial n$ and CPT results do lead to $\partial\mu/\partial n$ smaller on
the electron- than on the hole-doped side.\cite{Senechal:2004} In addition, the optical gap at half-filling is smaller in electron- than in hole-doped systems.

What is the physics? The physics of the antiferromagnetic phase at both weak
and strong coupling is well understood and needs no further comment. For the
pseudogap, we have argued that there seems to be two mechanisms, a weak
coupling one that involves scattering off critical fluctuations and that is
very well understood within TPSC, and a strong-coupling one where there is no
need for large correlation lengths. There is no simple physical picture for
the latter mechanism although the fact that it does not scale
with $J$ but with $t$ seems to suggest forbidden hopping.
The pseudogap is clearly different from the Mott gap.
Whether there is a phase transition as a function of $U$ that separates the
weak and strong coupling regimes or whether there is only a crossover is an
open question. The shape of the MDC's at the Fermi energy clearly show in any
case that in some directions wave vector is not such a bad quantum number
whereas in the pseudogap direction, a \textquotedblleft
localized\textquotedblright\ or \textquotedblleft almost
localized\textquotedblright\ particle-like picture would be appropriate. In
fact the pseudogap occurs near the intersection with the antiferromagnetic
zone boundary that turns out to also be the place where umklapp processes are
possible. In other words, the presence of a lattice is extremely important for
the appearance of the pseudogap. We have seen that with spherical Fermi
surfaces the Fermi liquid survives even for large $U$. The dichotomy between
the wave description inherent to the Fermi liquid and the particle (localized)
description inherent to the Mott phenomenon seems to be resolved in the
pseudogap phase by having certain directions where electrons are more
wave-like and other directions where particle-like (gapped) behavior appears.
The latter behavior appears near regions where the presence of the lattice is
felt through umklapp processes.

It is clear that when weak-coupling-like ideas of quasiparticles scattering
off each other and off collective excitations do not apply, a simple physical
description becomes difficult. In fact, knowing the exact wave functions would
give us the solution but we would not know how to understand \textquotedblleft
physically\textquotedblright\ the results.

This lack of simple physical images and the necessity to develop a new
discourse is quite apparent for d-wave superconductivity. At weak coupling
exchange of slow antiferromagnetic fluctuations is at the origin of the
phenomenon, while at strong-coupling the fact that the maximum value of the
d-wave order parameter scales with $J$ tells us that this microscopic coupling
is important, even though there is no apparent boson exchange. This is where
mean-field like theories\cite{Kotliar:1988, Inui:1988, Anderson:2005} or
variational approaches\cite{Paramekanti:2004} can help when they turn out to
give results that are confirmed by more accurate and less biased methods.

There are many open problems, some of which are material dependent and hence
may depend on interactions not included in the simplest Hubbard model. We have
already mentioned the problem of the chemical potential shift in ARPES for
very small dopings\cite{Shen:2005} that seems to be somewhat material
dependent\cite{Yoshida:2005}. It would also be important to understand
additional inhomogeneous phases that are observed in certain high-temperature
superconductors. That is extremely challenging for quantum cluster methods and
unlikely to be possible in the very near future, except for inhomogeneities of
very short wave length. Also, we still need to improve concordance between the
methods before we can make predictions that are quantitative at the few
percent level for all physical quantities. Apart from DCA, there are no
quantum cluster methods that have been developed yet to study two-particle
response functions that are necessary to obtain results on the superfluid
density and on transport in general. Transport studies are being completed in
TPSC.\cite{Allen:unpub2} Such studies are crucial since they are needed to
answer questions such as: (i) Why is it that for transport properties, such as
optical conductivity, the number of carriers appears to scale with doping
whereas in ARPES the surface of the Brillouin zone enclosed by the apparent
Fermi surface appears to scale with the number of electrons? Is it because the
weight of quasiparticles at the Fermi surface scales like the doping or
because of vertex corrections or because of both? (ii) Can we explain a
vanishing superfluid density as doping goes to zero\cite{Broun:2005} only
through Mott physics or can competing order do the job.\footnote{I.~Herbut,
private communication},\cite{Micnas:2005}

After twenty years all the problems are not solved, but we think that we can
say with confidence that the essential physics of the problem of
high-temperature superconductivity is in the one-band Hubbard model. At least
the pseudogap, the antiferromagnetic and the d-wave superconducting phases
come out from the model. Refinements of that model may however be necessary as
we understand more and more details of the material-specific experimental results.

Has a revolution been necessary to understand the basic physics of
high-temperature superconductors? Certainly, it has been necessary to change
our attitude towards methods of solution. We have seen that to study
intermediate coupling, even starting from weak coupling, it has been necessary
to drop diagrams and to rely instead on sum rules and other exact results to
devise a non-perturbative approach. At strong coupling we had to accept that
numerical methods are essential for progress and that we need to abandon some
of the traditional physical explanations of the phenomena in terms of
elementary excitations. Even though progress has been relatively slow, the
pace is accelerating in the last few years and there is hope that in a few
years the problem will be considered for the most part solved. The theoretical
methods (numerical and analytical) that have been developed and that still
need to be developed will likely remain in the tool box of the theoretical
physicist and will probably be useful to understand and perhaps even design
other yet undiscovered materials with interesting properties. The success will
have been the result of the patient and focused effort of a large community of
scientists fascinated by the remarkable phenomenon of high-temperature superconductivity.

\begin{acknowledgments}
The present work was supported by NSERC (Canada), FQRNT (Qu\'{e}bec), CFI
(Canada), CIAR, the Tier I Canada Research chair Program (A.-M.S.T.). We are
grateful to our collaborators, G. Albinet, S.~Allen, M.~Boissonneault,
C.~Brillon, M.~Capone, L.~Chen, M.~Civelli, A.-M. Dar\'{e}, B.~Davoudi, J.-Y.
P. Delannoy, A.~Gagn\'e-Lebrun, M. J. P. Gingras, A.~Georges, M.~Guillot, V. Hankevych, P. C. W.
Holdsworth, F.~Jackson, S.~Kancharla, G.~Kotliar, J.-S.~Landry, P.-L.~Lavertu, F.~Lemay, S.~Lessard,
M.-A.~Marois, S.~Pairault, D.~Perez, M.~Pioro-Ladri\`{e}re, D.~Plouffe,
D.~Poulin, L. Raymond, S.~Roy, P.~Sahebsara, H.~Touchette, and especially
Y.M.~Vilk. We also acknowledge useful discussions with P.~Fournier, M. Greven,
I.~Herbut, K. Shen and L.~Taillefer and we are grateful to V. Hankevych and S.
Kancharla for permission to include some of their unpublished figures in this paper.
\end{acknowledgments}

\appendix

\section{List of acronyms}

\begin{description}
\item ARPES: Angle Resolved Photoemission Spectroscopy: Experiment from which one can extract $A(\mathbf{k,}\omega)f(\omega)$.

\item CPT: Cluster Perturbation Theory: Cluster method based on strong
coupling perturbation theory.\cite{Gros:1993, Senechal:2000,Senechal:2002}

\item CDMFT: Cellular Dynamical Mean Field Theory: A cluster generalization
of DMFT that allows one to take into account both wave vector and frequency
dependence of the self-energy.\cite{Kotliar:2001} It is best formulated in
real space.

\item DCA: Dynamical Cluster approximation: A cluster generalization of DMFT
that allows one to take into account both wave vector and frequency dependence
of the self-energy based on coarse graining of the self-energy in reciprocal
space.\cite{Hettler:1998, Hettler:2000}

\item DMFT: Dynamical Mean Field Theory: This approach is exact in infinite
dimension. It takes the frequency dependence of the self-energy into account
and includes both the Mott and the Fermi liquid
limits.\cite{Georges:1996,Jarrell:1992}

\item EDC: Energy Dispersion Curves: A representation of $A\left(
\mathbf{k,}\omega\right)  f\left(  \omega\right)  $ at fixed $\mathbf{k}$ as a
function of $\omega.$

\item FLEX: Fluctuation Exchange Approximation: A conserving many-body
approach, similar in spirit to Eliashberg theory.\cite{Bickers:1989}

\item MDC: Momentum Dispersion Curves: A representation of $A(
\mathbf{k,}\omega)f(\omega)$ at fixed $\omega$ as a
function of $\mathbf{k}$.

\item QMC: Quantum Monte Carlo: Determinental
approach\cite{Blankenbecler:1981}. This provides an essentially exact solution
to the model for a given system size and within statistical errors that can be
made smaller by performing more measurements.

\item RPA: Random Phase Approximation.

\item TPSC: Two-Particle Self-Consistent Approach: Based on sum rules and
other constraints, allows to treat the Hubbard model non-perturbatively in the
weak to intermediate coupling limit.\cite{Vilk:1997, Allen:2003}

\item VCA: Variational Cluster Approach. Analogous to \textit{CDMFT} but
with a convergence criterion based on an extremum principle. In the
applications quoted here, there is no bath.\cite{Potthoff:2003a,
Potthoff:2003b, Potthoff:2005, Dahnken:2004}

\item VCPT: Variational Cluster Perturbation Theory. In the present paper
synonymous with \textit{VCA.}

\end{description}


\end{document}